\documentclass[prb,reprint, amsmath,amssymb,showpacs,superscriptaddress,floatfix]{revtex4-2}
\usepackage[breaklinks=true,colorlinks,citecolor=blue,linkcolor=blue,urlcolor=blue]{hyperref}
\usepackage[table,xcdraw]{xcolor}

\usepackage{epsfig,mathrsfs,color,latexsym,subfigure,marginnote,gensymb,}
\usepackage{graphicx}
\usepackage{xcolor}
\usepackage{booktabs}
\renewcommand{\BibitemShut}[1]{}

\begin{document}
\title{Room Temperature Ferroelectricity and Electrically Tunable Berry Curvature Dipole in III-V Monolayers}
\author{Ateeb Naseer}
\affiliation{Department of Electrical Engineering, Indian Institute of Technology, Kanpur, Kanpur 208016, India}
\author{Achintya Priydarshi}
\affiliation{Department of Electrical Engineering, Indian Institute of Technology, Kanpur, Kanpur 208016, India}
\author{Pritam Ghosh}
\affiliation{Department of Materials Science \& Engineering, Indian Institute of Technology, Kanpur, Kanpur 208016, India}
\author{Raihan Ahammed}
\affiliation{Department of Physics, Indian Institute of Technology, Kanpur, Kanpur 208016, India}
\author{Yogesh Singh Chauhan}
\affiliation{Department of Electrical Engineering, Indian Institute of Technology, Kanpur, Kanpur 208016, India}	
\author{Somnath Bhowmick}
\email{bsomnath@iitk.ac.in}
\affiliation{Department of Materials Science \& Engineering, Indian Institute of Technology, Kanpur, Kanpur 208016, India}
\author{Amit Agarwal}
\email{amitag@iitk.ac.in}
\affiliation{Department of Physics, Indian Institute of Technology, Kanpur, Kanpur 208016, India}
\date{\today}
\begin{abstract}
Two-dimensional ferroelectric monolayers are promising candidates for compact memory devices and flexible electronics.  Here, through first-principles calculations, we predict room temperature ferroelectricity in AB-type monolayers comprising group III (A = Al, In, Ga) and group V (B = As, P, Sb) elements. We show that their spontaneous polarization, oriented out-of-plane, ranges from 9.48 to 13.96 pC/m, outperforming most known 2D ferroelectric. We demonstrate electric field tunable Berry curvature dipole and nonlinear Hall current in these monolayers. Additionally, we highlight their applicability in next-generation memory devices by forming efficient ferroelectric tunnel junctions, especially in InP, which supports high tunneling electroresistance. Our findings motivate further exploration of these monolayers for studying the interplay between Berry curvature and ferroelectricity and for integrating these ferroelectric monolayers in next-generation electronic devices.  
\end{abstract}
\maketitle
\section{Introduction}
Ferroelectric materials are pivotal for several high-speed, low-power applications such as field-effect transistors, high-density memory devices, and sensors~\cite{zhuravlev_giant_2005,wang_direct_2020,yan_high-performance_2011,Musaib_FE,wang_field-effect_2014, Achintya, Musaib_FE1, hoffmann_unveiling_2019}. The recent push towards miniaturization in device engineering has shifted the focus to two-dimensional (2D) ferroelectric materials and thin films. Thin films have a problem, as reducing material thickness leads to a decline in electric polarization due to unscreened depolarization fields~\cite{dep_1,dep_2,dep_3,dep_4,dep_5}. Owing to this, the discovery and study of intrinsic 2D ferroelectric materials becomes crucial~\cite{Chhowalla2016,Das2021_2D,Nandan2023, Nov, Nov1, Ateeb_1,Yadav_TMDs, Yoon, Kou, Rodin, Ateeb2,belianinov_cuinp2s6_2015,liu_room-temperature_2016, chang_discovery_2016, cui_intercorrelated_2018, xiao_intrinsic_2018,wan_room-temperature_2018,poh_molecular-beam_2018,fei_ferroelectric_2018,zheng_room_2018, PhysRevApplied.13.044014, PhysRevB.103.075436}. Beyond their ferroelectric properties, 2D materials offer atomic-scale thickness and pristine interfaces for integration with flexible and low-dimensional electronic and memory devices. A promising application of 2D ferroelectrics is in ferroelectric tunnel junctions (FTJs). The monolayer limit of ferroelectric layers helps achieve high tunnel electroresistance (TER) with reduced size. Studies have shown remarkable TER performance in 2D FTJs, rivaling their 3D counterparts~\cite{kang_realizing_2019, shen_two-dimensional_2019}. 

Since materials supporting ferroelectricity always lack an inversion center, enabling a finite Berry curvature dipole (BCD), giving rise to the intriguing phenomena of nonlinear anomalous Hall effect~\cite{PhysRevLett.115.216806, PhysRevB.92.235447, Subhajit_NPhy_23,Chakraborty_2022}. Combined with the fact that the electronic properties of 2D materials can be tuned by applying a vertical electric field, this makes 2D ferroelectric materials an electrically tunable platform for novel nonlinear optoelectronic and transport responses. The band topology of 2D ferroelectrics can also be tuned electrically, offering a testbed for studying the interplay of electrically controlled ferroelectricity and Berry curvature~\cite{PhysRevLett.49.405, RevModPhys.82.1959, science.1187485, science.1089408,Wang_2019,PhysRevB.102.024109}. Interestingly, earlier studies have demonstrated the possibility of manipulating photocurrents and nonlinear Hall signals in materials like SnTe and WTe$_2$, owing to their significant BCD~\cite{kim2019prediction, isobe2020high, ma2019observation, kang2019nonlinear, xu2018electrically}. Such discoveries highlight the potential of these materials for designing devices for frequency doubling, rectification, wireless communications, and energy harvesting. 

Here, we predict group III-V monolayers (generic formula AB; A = Al, Ga, In and B = P, As, Sb) to be ferroelectric at room temperature. We show that they exhibit out-of-plane spontaneous polarization, surpassing most known 2D ferroelectrics in terms of the magnitude of polarization. Their relatively high dipole-dipole interaction strength stabilizes their ferroelectricity at high temperatures, enhancing their practical applicability. Among these, InP stands out for FTJ-based memory applications, with TER comparable to BaTiO$_3$ thin films. More interestingly, the synergy of ferroelectricity and Berry curvature in these materials endows them with electric field tunable optoelectronic and nonlinear transport properties. Our comprehensive study highlights the potential of these novel III-V monolayers for ferroelectric tunnel junctions, memory devices, and nonlinear optoelectronics and for studying the interplay between band topology and ferroelectricity.  

\begin{figure*}
\includegraphics[width=\textwidth]{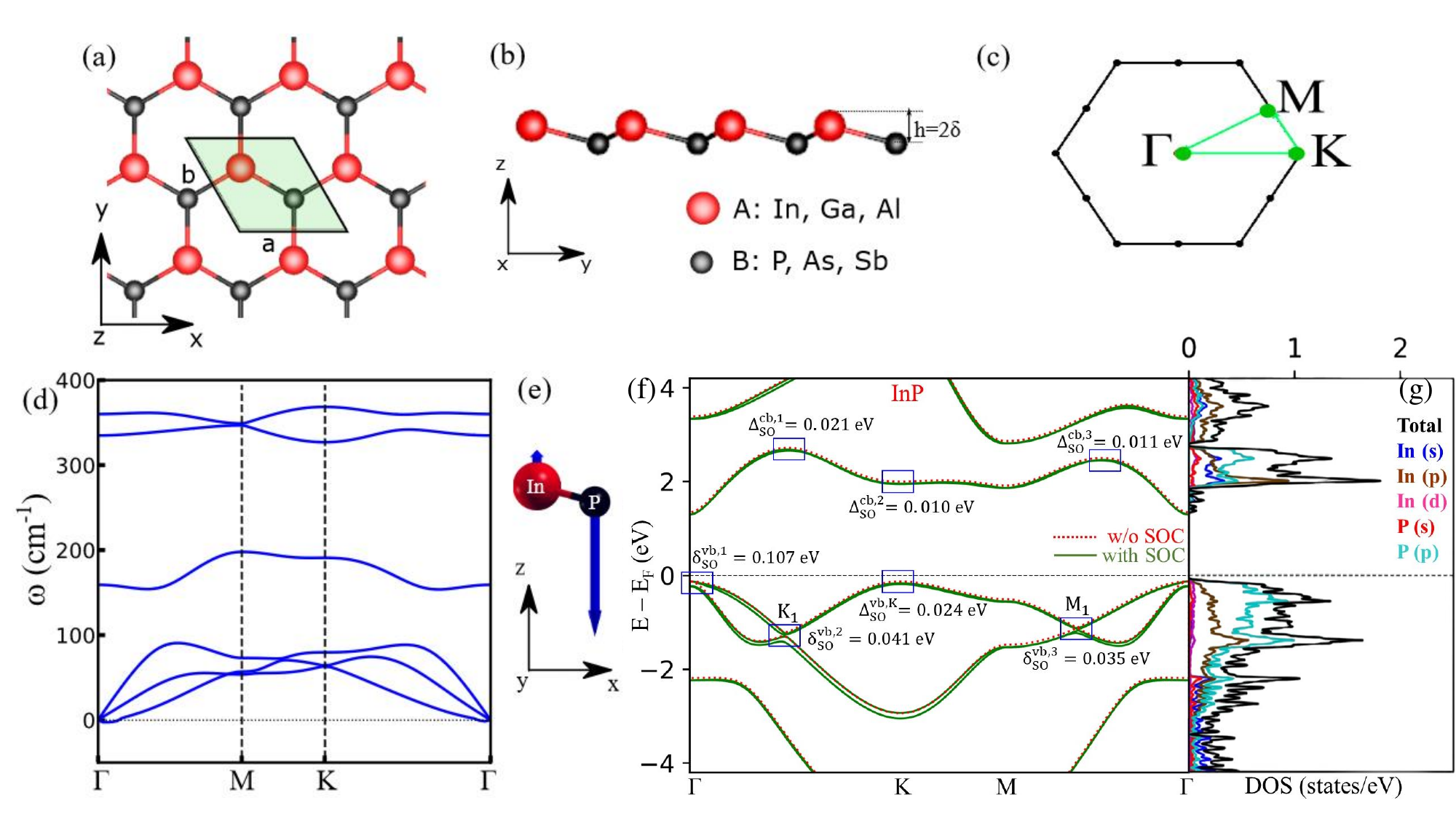}
\caption{(a) Top view and (b) side view of AB monolayer, with buckling height $h=2\delta$. (c) Hexagonal Brillouin zone and high symmetry points. (d) Phonon dispersion curves and the (e) polar vibrational mode for the InP monolayer. The blue arrows show the direction of the atomic displacement. (f) The electronic bandstructure of InP monolayer with SOC (green solid line) and without SOC (red dotted line) and (g) orbital projected density of states, showing the orbital contribution of each atom for InP. The p-orbital of P and In atoms have the most significant contribution near the valence and conduction band edges.}
\label{fig:Fig1}
\end{figure*}

\begin{figure*} 
\includegraphics[width=\textwidth]{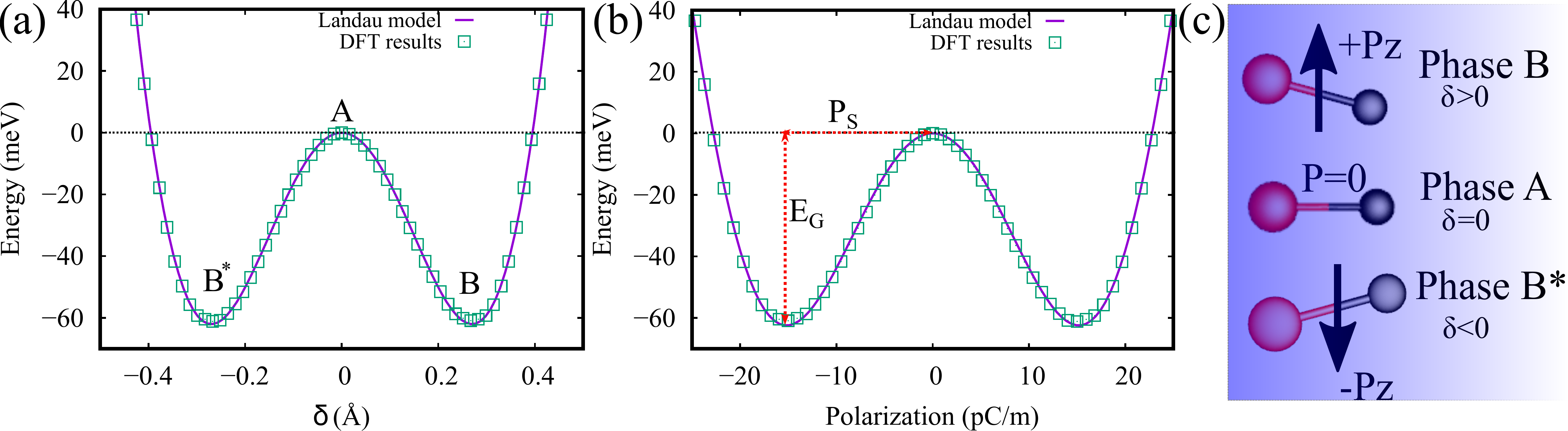}
\caption{Double well type variation of the energy for InP, (a) as a function of $\delta$ [see Fig.~\ref{fig:Fig1}(b)], and (b) as a function of polarization. The spontaneous polarization and energy barrier for ferroelectric phase transition are P$_s$ and E$_G$, respectively. (c) The transition between non-centrosymmetric ferroelectric phases with non-zero polarization via a centrosymmetric paraelectric state with zero polarization.}
\label{phonon}
\end{figure*}
\begin{table*}
\label{tab:1}
\caption{The parameters for the ferroelectric phase of different monolayers. The potential barrier $E_G$ (meV), spontaneous polarization $P_{s}$ (10$^{-12}$ C/m), Curie temperature ($T_C$) in $K$, coercive field, $E_c$ (V/nm), LG fitting parameters in Eq.(1), $A$ [meV/(pC/m)$^2$], $B$ [meV/(pC/m)$^4$] (10$^{-2}$), $C$ [meV/(pC/m)$^6$] (10$^{-5}$), describing the double-well potential of the energy vs. polarization plots, and $D$ [meV/(pC/m)$^2$], describing the average dipole-dipole interaction.}
\centering
\begin{tabular}{@{}p{0.1\linewidth}p{0.1\linewidth}p{0.1\linewidth}p{0.1\linewidth}p{0.1\linewidth}p{0.1\linewidth}p{0.1\linewidth}p{0.1\linewidth}p{0.1\linewidth}@{}}
\toprule
            Material    & $E_{G}$ & $P_{s}$ & $T_C$ &$E_{c}$ & $A$ & $B$ & $C$ & $D$ \\
            \midrule
            AlSb &  102.37  &  9.48 & 5207   &  1.41    & -4.56 & 5.80 &-12.06 & 4.99   \\
            GaAs &  129.86  &  9.87 & 5124   &  1.95  & -5.20 & 5.32 &-9.18 & 4.53       \\
            GaP  &   28.94 &  10.18 & 1711   &  0.44    &-1.02 & 0.88 &-0.54 & 1.42      \\
            InAs &  172.99  & 13.67 & 6517   &  1.56  & -4.18 & 3.16 &-5.16 & 3.00     \\
            InP  &   61.36 &  13.96 & 3284  &   0.55    & -1.16 & 0.60 &-0.48 & 1.45    \\
            InSb &   218.07 & 10.06 & 7986 &   2.47    & -8.56 & 10.12 &-24.18 & 6.80  \\
			\bottomrule
\end{tabular}
\end{table*}
\section{Computational details}
We perform \textit{ab initio} calculations using density functional theory (DFT) and projector-augmented wave pseudo-potentials, as implemented in the Quantum Espresso (QE) package \cite{giannozzi_quantum_2009,giannozzi_advanced_2017}. To calculate the exchange and correlation energies, we use the Perdew-Burke-Ernzerhof (PBE)~\cite{blochl_projector_1994} implementation of the generalized gradient (GGA) approximation. The energy cutoff for the plane wave basis set is 50 Ry. We use a $k$-mesh of $24\times24\times1$ for sampling the Brillouin zone. Structural relaxations use force and energy convergence threshold of $10^{-3}$ Ry/a.u and $10^{-4}$ Ry, respectively. Grimme's DFT-D method describes the effects of the vdW interactions. To eliminate the spurious interactions between neighboring slabs, a vacuum layer of $\sim$ 20~\AA~is inserted between adjacent layers along the $z-$direction (perpendicular to the plane of monolayers). We used VESTA and XCRYSDEN for visualizing the crystal structures and phonon modes, respectively~\cite{momma_vesta_2011, XCRYSDEN}. To determine the spontaneous polarisation ($P_{s}$) of III-V monolayers, we use the Berry phase method~\cite{Berry_1,Berry_2,Berry_3}.

\section{Results and Discussions}
\subsection{Crystal structure, Dynamical stability and electronic properties} We illustrate the top and side view of group III-V monolayers (AB: A = Al, Ga, Sb: B = P, As, Sb) in Fig.~\ref{fig:Fig1} (a) and (b). Their buckled honeycomb structure is formed by adding a two-atom motif  (A and B) to the hexagonal lattice. Atoms A and B are located in two different planes, resulting in two distinct sublattices, each having a three-fold rotation axis. The optimized lattice constants (denoted by $a=b$) and buckling heights between two planes (denoted by $h=2\delta$), are in excellent agreement with the previously reported values~\cite{di_sante_emergence_2015,sahin_monolayer_2009}. This validates our choice of computational parameters described in the computational details section. We summarize the structural parameters in Table S1 of the Supplementary Information (SI). Calculated values of formation energies (using free atoms as reference) range from -3.04 to -4.05 eV/atom, indicating stability of the monolayers. Cleavage energy ranges from 0.13 to 0.34 J/m$^2$, which is comparable to graphene (0.32 J/m$^2$). Experimentally, ultra-thin III-V semiconductors have been fabricated using layer-by-layer growth on suitable substrates~\cite{Chen2020UniversalGO}.

\begin{figure*}
\includegraphics[width=\textwidth]{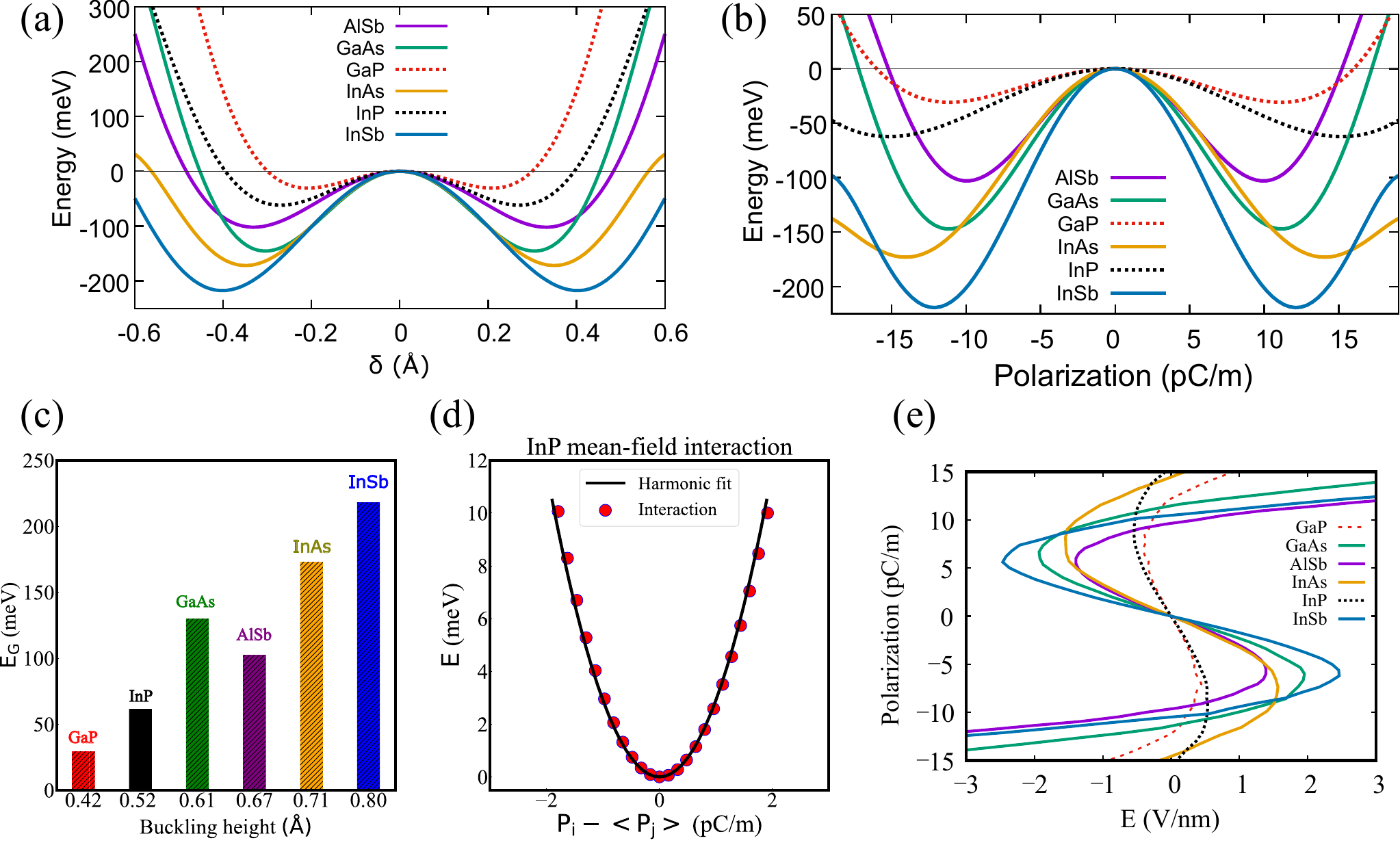}
\caption{The variation of (a) energy with $\mathrm{\delta}$ [see Fig.~\ref{fig:Fig1}(b)] and (b) energy with polarization. (c) The variation of the polarization switching barrier with buckling height for all monolayers depicts an almost linear variation. (d) The dipole-dipole interaction of monolayer InP, calculated using mean-field theory. (e) The variation of polarization with an electric field for all monolayers. This S-curve was obtained from the fitting of the Landau polynomial to DFT data, as shown in Fig.~\ref{phonon}(b).}  
\label{fig:Fig3}
\end{figure*}
 
We analyze the phonon spectra of the freestanding buckled monolayers to confirm their dynamic stability. We find that there are three acoustic and three optical modes. The absence of imaginary frequencies in the phonon dispersions indicates their structural stability. Our phonon dispersion calculations also reveal the phonon modes responsible for the out-of-plane spontaneous polarization of the III-V monolayers [Fig.~\ref{fig:Fig1} (d), (e) and Fig. S1, SI]. 

We present the electronic bandstructure of the InP monolayer in Fig.~\ref{fig:Fig1} (f),  along the high symmetry path of $\mathrm{\Gamma}$-K-M-$\mathrm{\Gamma}$ of the Brillouin zone, as shown in Fig.~\ref{fig:Fig1} (c). Our calculation of the orbital projected density of states reveals that the valence band maximum (VBM) and conduction band minimum (CBM) states are dominated by the p-orbitals of the constituent atoms. The presence of spin-orbit coupling (SOC) leads to the splitting of the top two-fold degenerate valence bands as illustrated in Fig.~\ref{fig:Fig1} (f) and Fig. S2, SI. This splitting results in a gap of $\delta_{SO}^{vb,1}=$ 107 meV at the $\Gamma$ point and $\delta_{SO}^{vb,2} = 41$ meV ($\delta_{SO}^{vb,3}=35$ meV) at the K$_1$ (M$_1$) points along the high symmetry K–$\Gamma$ (M–$\Gamma$) paths. The SOC splitting of the top valence band is $\Delta_{SO}^{vb,K}=$ 24 meV at the K point. Similarly, the SOC splittings of the bottom conduction band are $\Delta_{SO}^{cb,1}=$ 21 meV, $\Delta_{SO}^{cb,2}=$ 10 meV, and $\Delta_{SO}^{cb,3}=$ 11 meV at three different k-points.  The top two valence bands in the high-symmetry K-$\Gamma$ (M-$\Gamma$) k-path mainly arise from S$_z$ (S$_y$) spin. However, the bottom conduction bands have all three spin components [see Fig. S3, SI]. We find that all the studied III-V AB monolayers are insulators, with most having a direct bandgap except for the GaP [see Fig. S4, SI].  

\subsection{Ferroelectricity in III-V monolayers}
Ferroelectric materials are characterized by spontaneous electric polarization that can be switched by applying an external electric field. Group III-V monolayers have two degenerate ferroelectric ground states (phase B and B$^*$) with equal but opposite $P_s$ values (Fig.~\ref{phonon}). Ferroelectric phase has a buckled crystal structure (Fig.~\ref{fig:Fig1}), and it belongs to the P3m1 space group, having no inversion symmetry~\cite{di_sante_emergence_2015}. The buckled crystal structure gives rise to spontaneous polarization in the out-of-plane direction. The polarization vector points up (phase B) or down (phase B$^*$), depending on the location of the group III and V atom about the plane lying in the middle [Fig.~\ref{phonon}(c)]. The transition from one non-centrosymmetric ferroelectric state B to another non-centrosymmetric ferroelectric state B$^{*}$ occurs via a centrosymmetric paraelectric state (A), having all the atoms lying in a common plane [Fig.~\ref{phonon}(c)]. Such a transition causes the out-of-plane polarization to shift from $P_s$ (phase B) to 0 (phase A) to $-P_s$ (phase B$^*$). The transition from phase B-A-B$^{*}$ can be physically realized by applying an external electric field in the vertical direction, paving the way for ferroelectric device applications.

Plotted as a function of $\delta$ (0.5$\times$buckling height) and polarization [Fig.~\ref{phonon} (a)-(b), Fig. S5, Fig. S6 in SI], free energy shows a characteristic double-well potential shape for ferroelectric materials. Two degenerate minimum points on the curve represent the ferroelectric ground states B and B$^*$, while the hump in the center corresponds to the paraelectric state A. The gap between the hump and energy minimum denotes the barrier to ferroelectric switching ($E_G$). Interestingly, the value of $E_G$ is closely related to the buckling height of III-V monolayers [Fig.~\ref{fig:Fig3} (c)]. Among the known 2D ferroelectric materials, $\alpha$-In$_2$Se$_3$ is reported to have a very high $E_G$ $\sim$ 1090 meV while the As, an elemental ferroelectric, is reported to have a very low $E_G$ $\sim$ 5.83 meV \cite{Val7, Curie_2}. In comparison, the III-V family has low to moderate $E_G$, ranging from 28.94 meV in GaP to 218.07 meV in InSb [Table I and Fig.~\ref{fig:Fig3}]. A higher magnitude of $E_G$ implies a more significant driving force for switching and phase transition. Spontaneous polarization ($P_s$) values of III-V monolayers range from 9.482 pC/m (AlSb) to 13.96 pC/m (InP) [Table I]. In terms of $P_s$ values, III-V monolayers outperform most reported materials [Table S2, SI]. The high polarization values makes them good candidates for memory applications based on Fe-FETs, as distinguishing between the two states becomes easier. We find that III-V monolayers have low Born effective charge ($Z^*_{zz} < 0.15e$), which is unlike cubic perovskites ($Z^* \sim 5e$).  Because of the small Born effective charge,  the III-V monolayers are expected to experience a weaker depolarization field and, thus, are good candidates for ultrathin ferroelectric applications~\cite{di_sante_emergence_2015}.

\subsection{Curie temperature} 
Having demonstrated the existence of ferroelectricity in III-V monolayers, we now focus on their transition temperature. To this end, the free energy ($G$, in units of meV per unit cell) can be expressed in terms of polarization ($P$, in units of pC/m) as ~\cite{LG_1}
\begin{equation}
\label{eq:1}
G=\sum_i \left[\frac{A}{2}P_i^2+\frac{B}{4} P_i^4+\frac{C}{6} P_i^6 \right]+\frac{D}{2} \sum_{\langle i, j\rangle}\left(P_i-P_j\right)^2~.
\end{equation}
Here, the first term is the Landau-Ginzburg (LG) free energy, and the last term describes the nearest dipole-dipole interactions, determined by the polarization difference between neighboring unit cells [Fig.~\ref{fig:Fig3} (d)]~\cite{Curie_1}. We calculate the coefficients A, B, and C by fitting the first term of equation~\ref{eq:1} with the free energy and polarization values obtained from our first principle calculations [see Fig.~\ref{phonon}(b) and Fig. S6 in SI]. We summarize the values of all fitting parameters in Table I.  

The dipole-dipole interactions mainly determine the Curie temperature ($T_C$) of the paraelectric to ferroelectric phase transition. A large value of dipole-dipole interactions (D) indicates resistance to thermal fluctuations, resulting in higher $T_C$ values~\cite{Curie_1,Curie_2}. We estimate the transition temperature using $K_BT_C\sim D\times P_s{^2}$, with $K_B$ being the Boltzmann constant in eV~\cite{Curie_1}. We find that all the III-V monolayers have $T_C$ values greater than 1000 K [Table I], comparable to other commonly known 2D ferroelectrics like SnSe, SnS, As,~\cite{Curie_1,Curie_2} among others. 

The method employed to estimate the Curie temperature generally overestimates the transition temperature, and the structure can possibly disintegrate at a much lower temperature~\cite{PhysRevB.97.024110, C7NR09006D}. We perform the \textit{ab initio} molecular dynamic (AIMD) simulations to check the structural stability of the InP monolayer, as it is the most promising for applications in ferroelectric tunnel junctions. The total run time for MD simulations is 2 ps, with a time step of 0.0007 ps. We fix the system's temperature using the Nosé thermostat. The monolayer is stable up to $\sim 600$ K and retains around $ 1/4$ of the polarization value calculated at 0 K. Thus, InP monolayer can be used for ferroelectric device applications at room temperature and above. Further details about the AIMD simulations are given in Fig. S7, SI.

\begin{figure*} 
\includegraphics[width=0.9\textwidth]{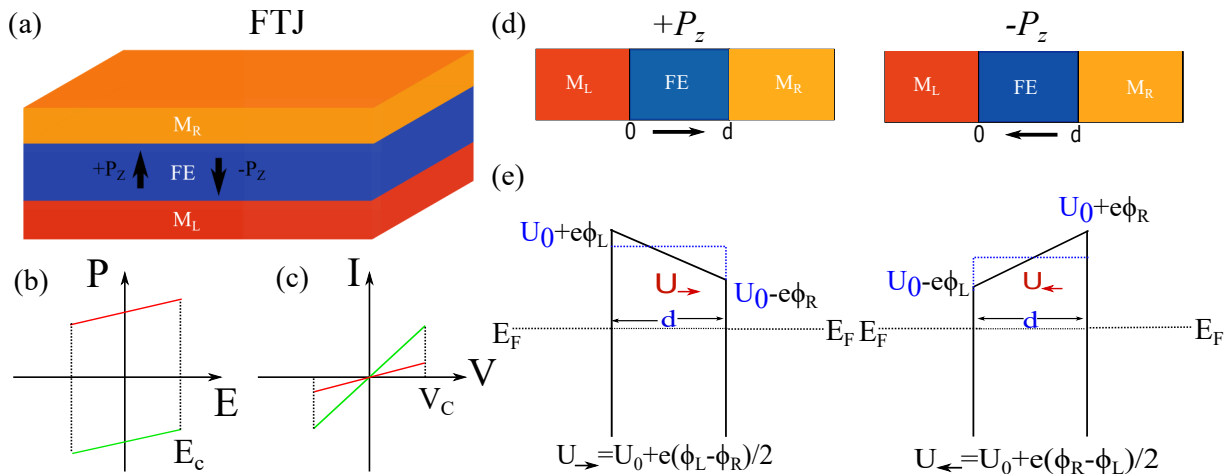}
\caption{(a) Schematic of a ferroelectric tunnel junction (FTJ) device consisting of a ferroelectric barrier between two metal electrodes. (b,c) Tunneling electroresistance (TER) effect shows the correlation between polarization and resistance of the FTJ. (d) Schematic of the device in  P$_{\rightarrow}$ and  P$_{\leftarrow}$ states. (e) Band diagram and potential energy profile of the device.}  
\label{fig:device_01}
\end{figure*}

\begin{figure}[b]
\includegraphics[width=.9\columnwidth]{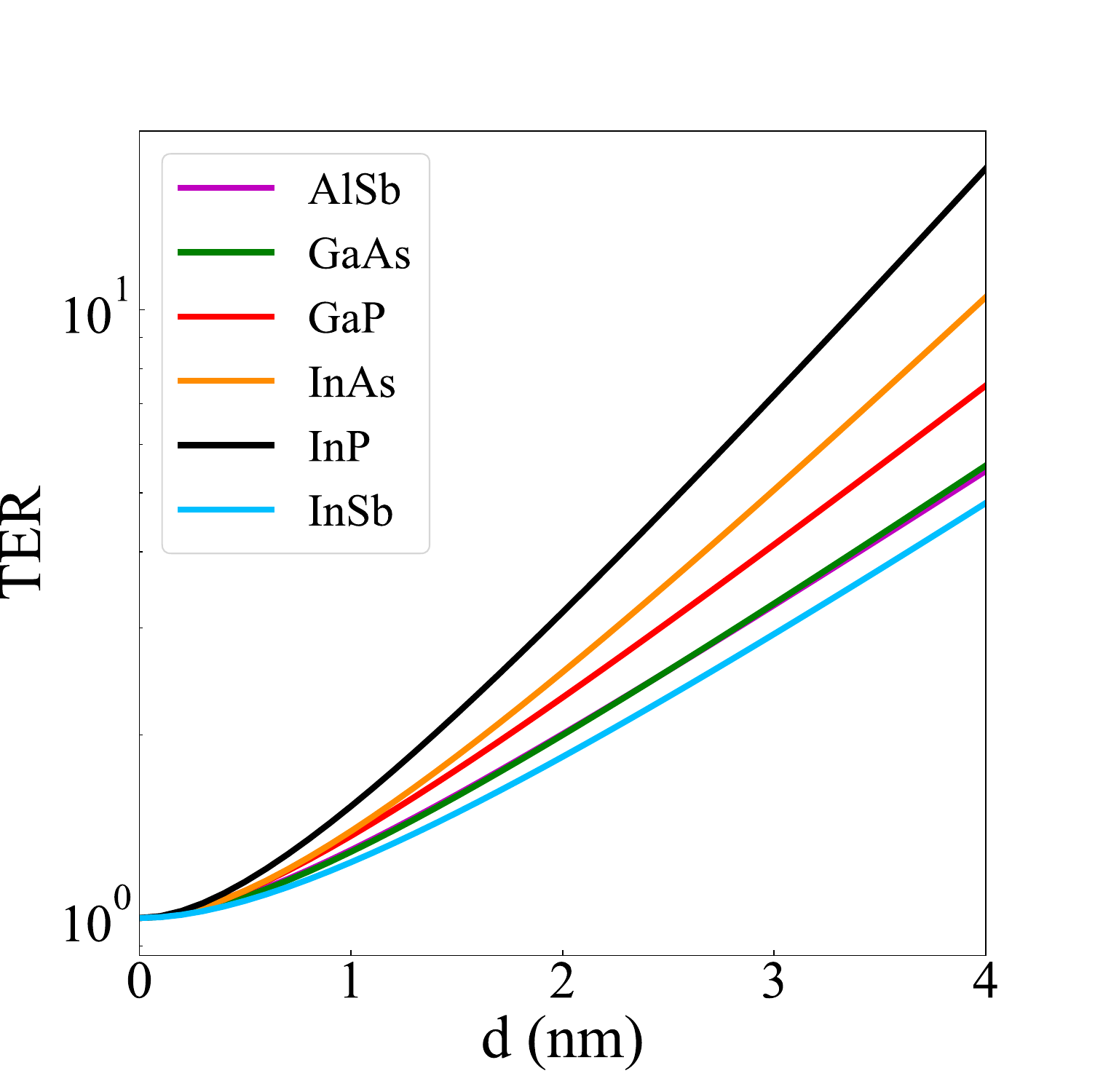}
\caption{The variation of the tunneling electroresistance of group III-V monolayers with the thickness of the ferroelectric layer (d).}
\label{fig:TER}
\end{figure}

\subsection{S-curve and estimating the coercive field} 
To estimate the coercive field of the III-V monolayers, we need to determine the  dependence of polarization on the electric field. For this, we use the relation dG/dP=E along with the Landau-Ginzburg free energy [first term of Eq.~\ref{eq:1}] to obtain the characteristic S-curve of ferroelectric materials. We present the calculated S-curves of III-V monolayers in Fig.~\ref{fig:Fig3}(e). The coercive field ($E_{c}$) is extracted from the turning points of the S-shaped curves using,
\begin{equation}
\left(\frac{dE}{dP}\right)_{E = E_{c}} = 0~. 
\end{equation}  
We find relatively low $E_{c}$ values for GaP ($\sim$ 0.441 V/nm) and InP ($\sim$ 0.546 V/nm), which are comparable to the most commonly used bulk ferroelectric, HfO$_2$ (generally between $\sim$ 0.1-0.3 V/nm~\cite{Wang_HfO2}). Rest of the monolayers have $E_{c}$ values larger than 1 V/nm, InSb being the largest ($\sim$ 2.5 V/nm). The optimum value of $E_{c}$ is desired for the memory device applications. The low value of $E_{c}$ will correspond to a small memory window, while a high $E_{c}$ will correspond to a larger memory window, but at the same time, it will require a large vertical field to switch from one state to other.

\begin{figure*}
\includegraphics[width=\textwidth]{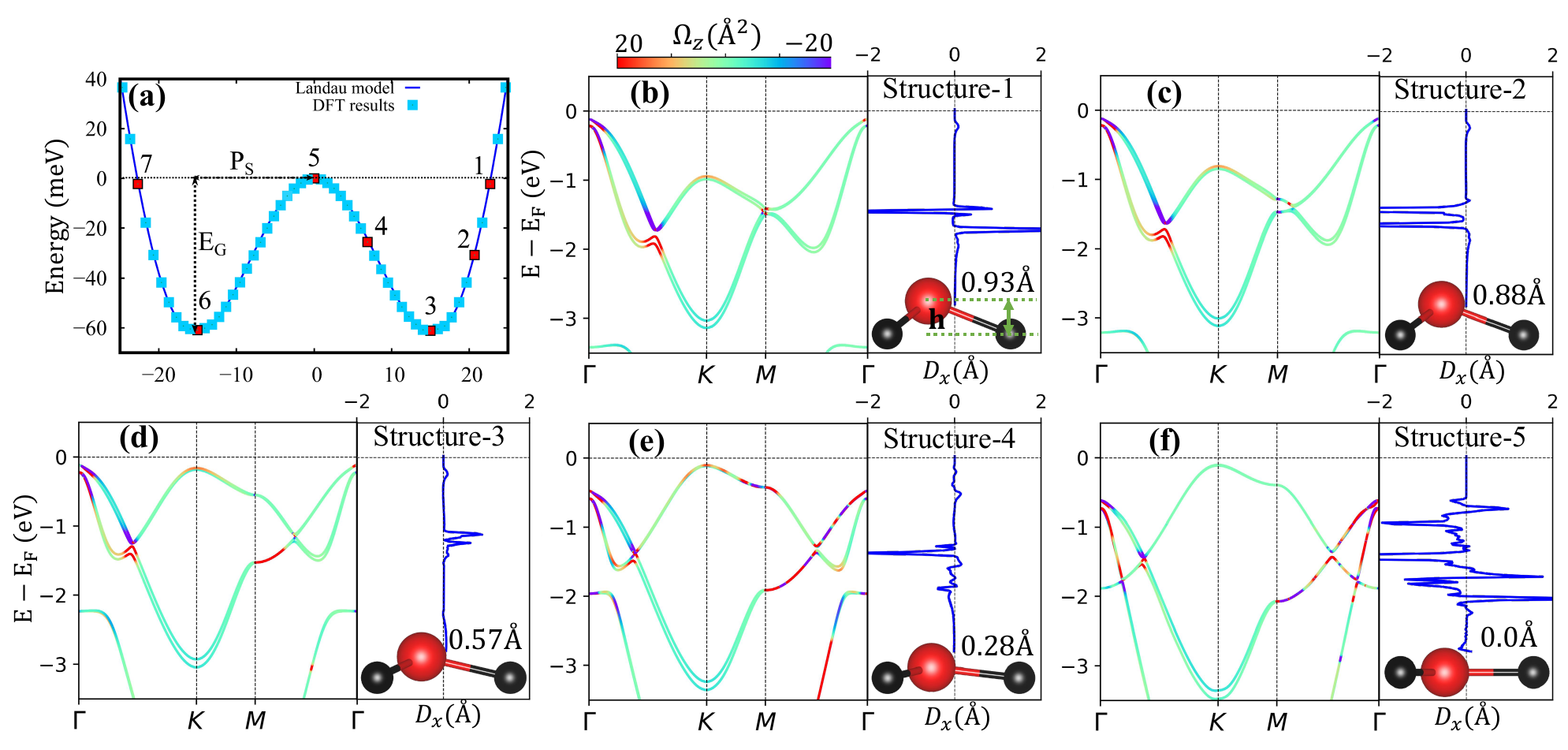}
\caption{(a) Double well shaped energy vs. polarization plot for InP monolayer. (b)-(f) Band structure and Berry curvature (BC) dipole (at $T = 100$ K) along x-axis of InP monolayer for various structures with different polarization. The color map represents the BC of the electronic bands. 
}
\label{fig:Band_BCD}
\end{figure*}

\begin{figure*}[t]
\includegraphics[width=\textwidth]{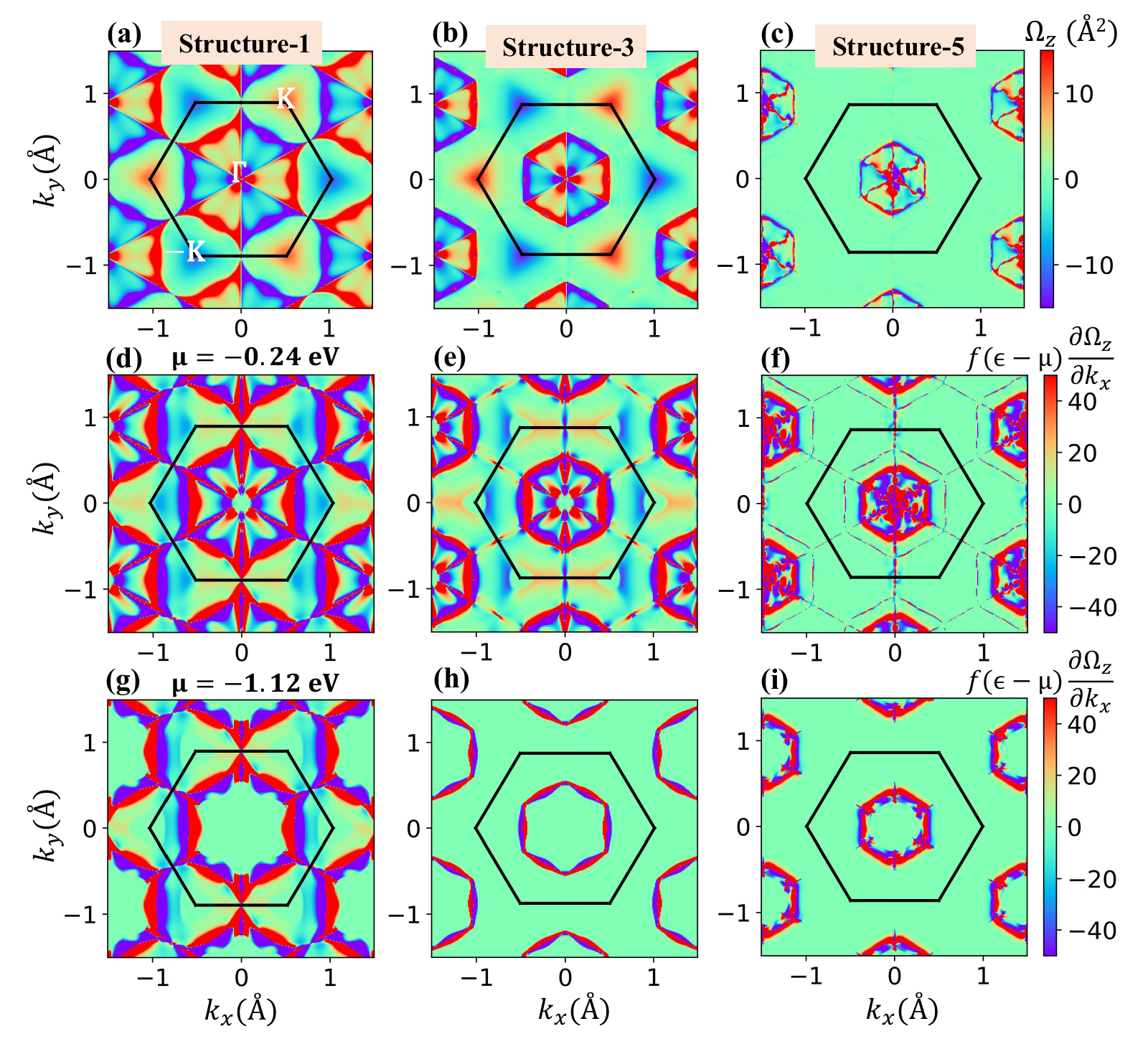}
\caption{For structure-1, structure-3 and structure-5 [see Fig.~\ref{fig:Band_BCD}(a)], (a)-(c) Berry curvature of the top valence band of InP monolayer with different polarizations; (d)-(f) BCD density with $\mu = -0.24 $ eV; (g)-(i) BCD density with $\mu = -1.12 $ eV. This highlights that changing the polarization state by applying a vertical electric field in the monolayer offers a knob to manipulate the BC distribution and the BCD even in monolayers.} 
\label{fig:BC_BCD}
\end{figure*}

\subsection{Ferroelectric Tunnel junction}
A ferroelectric tunnel junction (FTJ) consists of an ultrathin ferroelectric layer between two metal electrodes, as shown in Fig.~\ref{fig:device_01} (a). The ferroelectric layer must be very thin to let the electrons tunnel through. However, obtaining a high-quality ultrathin layer of conventional  ferroelectric materials is difficult. In this regard, 2D materials like III-V monolayers hold promise as they can sustain ferroelectricity down to a single layer. In FTJ, electrons near the ferroelectric/electrode interface try to screen the polarization charges arising from the spontaneous polarization of the FE layer. Incomplete screening gives rise to an additional electrostatic potential at the interface, given by
 \begin{equation}   \label{phi}
	\phi_{L/R}=\pm\dfrac{\Gamma_{L/R}P_sd}{d+\epsilon_{FE}(\Gamma_{L}+\Gamma_{R})}~.
\end{equation}
Here, $\Gamma_{L/R}(=\delta_{L/R}/\epsilon_{L/R}$) is the ratio of the left/right electrode screening length and dielectric permittivity. In Eq.~\eqref{phi}, $P_s$ denotes the induced charge density at the ferroelectric/electrode interface, while $d$ and $\epsilon_{FE}$ denote the height and dielectric constant of the ferroelectric layer, respectively.

As shown in Fig.~\ref{fig:device_01} (e), the potential profile takes the shape of a trapezoid instead of a fixed rectangular barrier of height $U_0$ (represented by the blue dotted line). We can tune the tunnel barrier by switching the polarization of the FE layer, with $U_{\rightarrow}$ $=$ $U_{0}$ + $e$($\phi_{L}$ - $\phi_{R}$ )/$2$ corresponding to $P_{\rightarrow}$ while $U_{\leftarrow}$ $=$ $U_{0}$ + $e$($\phi_{R}$ - $\phi_{L}$ )/$2$ corresponding to $P_{\mathrm{\leftarrow}}$ [see Fig.~\ref{fig:device_01}(e)]. By choosing two metals having different screening lengths, we can ensure that the barrier height is higher (OFF state) when polarization is along $P_{\rightarrow}$ and lower (ON state) when polarization is along $P_{\mathrm{\leftarrow}}$ [Fig.~\ref{fig:device_01}]. 

We approximate the TER using the WKB approximation in the linear region, $\Delta$ $U$ ($=$ $U_{\rightarrow}$ - $U_{\leftarrow}$) $\ll$ $U_{0}$. In this approximation, the TER ratio is given by ~\cite{TER_form_1,TER_form_2,TER_form_3}:  
\begin{equation} \label{eq:3}
{\rm TER}\approx \left[ \dfrac{\sqrt{2m}\Delta U}{\hbar \sqrt{U_{0}}}\right] = \exp \left[\dfrac{e}{\hbar}\sqrt{\dfrac{2m}{U_0}} \dfrac{P_{s}\left(\Gamma_{L}-\Gamma_{R} \right)d^{2} }{d+\epsilon_{FE}(\Gamma_{L}+\Gamma_{R})}\right].
\end{equation}    
Here, m is the carrier effective mass. Direct tunneling is the governing mechanism for the thin ferroelectrics at low bias. Since effective potential barrier height determines the electroresistance, the WKB method is a good first-order approximation to describe the tunneling current. Results obtained from the WKB approximation are found to be in good agreement with the experimental observations~\cite{TER_form_3}. We tabulate all the parameters associated with eq.~\ref{eq:3} in the SI (Table S3). We find that among the III-V monolayers under investigation, InP has the highest TER ratio and is more suitable for nanoscale TER-based memory devices. The thin film of well studied ferroelectric BaTiO$_3$ of  thickness 2.4 nm has a TER of 7, whereas InP in our study has a TER of 4.4 for the same thickness~\cite{TER_form_2}. Interestingly, the TER can be increased further by changing the thickness d, as shown in Fig.~\ref{fig:TER}.

\subsection{Electrically tunable Berry curvature dipole} 
Having demonstrated ferroelectric tunnel junction in III-V monolayers, we now focus on another of its exciting physical property, its Berry curvature dipole. The Berry curvature (BC) of a quantum material significantly influences its transport properties, acting as an effective magnetic field in momentum space which modifies the electron motion in real space \cite{Lahiri_22,PhysRevLett.129.227401}. The BC emerges under symmetry-broken environments (either time-reversal symmetry or inversion-symmetry has to be broken). It is one manifestation of the quantum geometry of Bloch wavefunctions of electronic bands, giving rise to an array of novel transport and optical phenomena \cite{PhysRevB.107.165131}. We find that all our studied III-V monolayers host a finite Berry curvature. However, the total Chern number for each band (and consequently the linear anomalous Hall effect) of all the monolayers has to vanish due to the material's time-reversal symmetry. In such inversion symmetry broken materials, the second-order nonlinear anomalous Hall response induced by the Berry curvature dipole (BCD) is the dominant Hall response without a magnetic field \cite{PhysRevB.108.L201405}. Terahertz detection can be achieved experimentally based on the intrinsic nonlinear Hall effect (NLHE) due to non-vanishing BCD in these ferroelectric 2D monolayers. Terahertz-band communication technology holds promise for future wireless networks, yet achieving rapid, sensitive, and broadband terahertz detection at room temperature remains challenging. The nonlinear Hall effect offers a solution by enabling direct terahertz radiation detection without forming semiconductor junctions. Such Hall rectifiers, leveraging the non-vanishing BCD-induced NLHE, provide a simple device architecture suitable for easy fabrication, integration on-chip, and potential mass production. Zhang et al. proposed a method for terahertz photodetection using NLHE in inversion asymmetric time-reversal invariant ferroelectric quantum materials ~\cite{zhang2021terahertz}. 

To illustrate the BCD-induced nonlinear Hall response in 2D ferroelectric III-V monolayers, we have chosen the InP monolayer as the representative system. The BCD is given as a function of chemical potential $\mu$ as follows
\begin{equation}
    D_\alpha (\mu)  = \frac{1}{(2\pi)^2}  \sum_n \int d{\bf k} f(\epsilon_{\bf k}^n - \mu) \frac{\partial \Omega_n({\bf k)}}{\partial k_\alpha}~.
    \label{eq:BCD}
\end{equation}%
Here, $\alpha$ denotes $(x, y)$, $\epsilon_{\bf k}^n$ is the energy of n-th band, and $f(\epsilon_{\bf k}^n - \mu)$ is the Fermi-Dirac distribution function. In Eq.~(\ref{eq:BCD}), a sum over all the occupied bands is implied. The Berry curvature $\Omega_n ({\bf k} )$ of the n-th band at \textbf{k}-point is calculated from the Bloch part of the first-principles wavefunction, using the Kubo formula given by
\begin{equation}
    \Omega_n ( {\bf k} ) = -2 \text{Im}\sum_{\substack{m\neq n}}\frac{\langle u_{\bf k}^n \vert v_x \vert u_{\bf k}^m \rangle \langle u_{\bf k}^m \vert v_y \vert u_{\bf k}^n \rangle }{\left(\epsilon_{\bf k}^n - \epsilon_{\bf k}^m \right) ^2}~.
    \label{eq:BC}
\end{equation}
Here, $v_{x,y}$ are the velocity operators, and $u_{\bf k}^n$ is the Bloch part of the wavefunction. We calculate the BC via maximally localized Wannier function computed through the VASP~\cite{PhysRevLett.77.3865, PhysRevB.54.11169} (Vienna \textit{ab initio} simulation package) suite of codes along with the WANNIER90~\cite{pizzi2020wannier90} package. 

We present the BCD of InP monolayer in Fig.~\ref{fig:Band_BCD} for various structures with different polarizations as marked in Fig.~\ref{fig:Band_BCD} (a). Experimentally, the polarization can be tuned by applying an external electric field. A finite BCD may appear due to doping or when the Fermi level is below the valence band maxima (VBM). We find that the BCD's value is significant around the energy levels of $E = - 1.2$ eV, where the spin-orbit coupling (SOC) opens up small gaps at the band-crossing point. For structure 3 of InP monolayer [as marked in Fig.~\ref{fig:Band_BCD}(a)] with $P_s = 13.96 $ pC/m, the calculated value of BCD along x-axis with $\mu = -1.12 $ eV is $D_x = 0.9$ \AA, which exceeds the values previously documented for SnTe monolayer (0.3 \AA)~\cite{kim2019prediction} and T$_d$ - WTe$_2$ ($\sim$ 0.1 - 0.3 \AA)~\cite{PhysRevB.98.121109, zhang2018electrically}. Using a vertical electric field to tune the distinct polarized phases of the InP monolayer, we can manipulate the BC dipole to control the nonlinear optoelectronic responses and transport properties. We present the distribution of the top valence band's BC of InP monolayer in Figs.~\ref{fig:BC_BCD}(a-c), and the corresponding BCD density is illustrated in Figs.~\ref{fig:BC_BCD}(d-i) with different chemical potentials. We find that the Berry curvature (BC) is zero at the alternate high-symmetric ($-$K/K) points for structure-5 as depicted in Fig.~\ref{fig:BC_BCD}(c). For structure-1 and structure-3, the BC exhibits opposite values at the $-$K/K points. Consequently, the valley Chern number is equal and opposite in the alternate valleys. But, the material's time-reversal symmetry eliminates the total Chern number for each band, which can be seen in Figs.~\ref{fig:BC_BCD}(a-c), where the sum of the BC is zero in the first BZ. However, the BCD density (Figs.~\ref{fig:BC_BCD}(d-i)) shows a non-uniform distribution over BZ for positive and negative values, which is visible in Fig.~\ref{fig:BC_BCD}(e). This results in a finite BCD, which generates second order anomalous Hall response.

\section{Conclusion} 
Our first-principles DFT calculations demonstrate that group III-V monolayers exhibit remarkable out-of-plane ferroelectricity at room temperature. These monolayers outperform most known 2D ferroelectric materials in spontaneous polarization. These monolayers, particularly InP with high polarization (13.96 pC/m) and moderate switching barrier (61.36 meV), are promising for next-generation ultrathin ferroelectric devices. InP's coercive field is comparable to bulk ferroelectrics like HfO$_2$, making it ideal for ferroelectric tunnel junction-based memory applications, matching the performance of BaTiO$_3$ thin films. Additionally, these monolayers hold significant potential for realizing a substantial Berry curvature dipole (BCD) and nonlinear Hall effect, which can be tuned electrically. Our findings motivate further exploration into these novel materials' physics and device applications, particularly in terms of their ferroelectricity and electrically tunable nonlinear transport and optical properties. 

\section*{ACKNOWLEDGMENTS}
We acknowledge financial support from the Prime Minister's Research Fellowship (PMRF) and Department of Science and Technology (DST), Government of India, under Grant DST/SJF/ETA-02/2017-18. We acknowledge the National Supercomputing Mission (NSM) for providing computing resources of PARAM SANGANAK at IIT Kanpur, which is implemented by C-DAC and supported by the Ministry of Electronics and information technology (MeitY) and Department of Science and Technology (DST), Government of India. We also acknowledge the HPC facility provided by CC, IIT Kanpur.

\bibliography{ref}

\begin{thebibliography}{83}%
\makeatletter
\providecommand \@ifxundefined [1]{%
 \@ifx{#1\undefined}
}%
\providecommand \@ifnum [1]{%
 \ifnum #1\expandafter \@firstoftwo
 \else \expandafter \@secondoftwo
 \fi
}%
\providecommand \@ifx [1]{%
 \ifx #1\expandafter \@firstoftwo
 \else \expandafter \@secondoftwo
 \fi
}%
\providecommand \natexlab [1]{#1}%
\providecommand \enquote  [1]{``#1''}%
\providecommand \bibnamefont  [1]{#1}%
\providecommand \bibfnamefont [1]{#1}%
\providecommand \citenamefont [1]{#1}%
\providecommand \href@noop [0]{\@secondoftwo}%
\providecommand \href [0]{\begingroup \@sanitize@url \@href}%
\providecommand \@href[1]{\@@startlink{#1}\@@href}%
\providecommand \@@href[1]{\endgroup#1\@@endlink}%
\providecommand \@sanitize@url [0]{\catcode `\\12\catcode `\$12\catcode
  `\&12\catcode `\#12\catcode `\^12\catcode `\_12\catcode `\%12\relax}%
\providecommand \@@startlink[1]{}%
\providecommand \@@endlink[0]{}%
\providecommand \url  [0]{\begingroup\@sanitize@url \@url }%
\providecommand \@url [1]{\endgroup\@href {#1}{\urlprefix }}%
\providecommand \urlprefix  [0]{URL }%
\providecommand \Eprint [0]{\href }%
\providecommand \doibase [0]{https://doi.org/}%
\providecommand \selectlanguage [0]{\@gobble}%
\providecommand \bibinfo  [0]{\@secondoftwo}%
\providecommand \bibfield  [0]{\@secondoftwo}%
\providecommand \translation [1]{[#1]}%
\providecommand \BibitemOpen [0]{}%
\providecommand \bibitemStop [0]{}%
\providecommand \bibitemNoStop [0]{.\EOS\space}%
\providecommand \EOS [0]{\spacefactor3000\relax}%
\providecommand \BibitemShut  [1]{\csname bibitem#1\endcsname}%
\let\auto@bib@innerbib\@empty
\bibitem [{\citenamefont {Zhuravlev}\ \emph {et~al.}(2005)\citenamefont
  {Zhuravlev}, \citenamefont {Sabirianov}, \citenamefont {Jaswal},\ and\
  \citenamefont {Tsymbal}}]{zhuravlev_giant_2005}%
  \BibitemOpen
  \bibfield  {author} {\bibinfo {author} {\bibfnamefont {M.~Y.}\ \bibnamefont
  {Zhuravlev}}, \bibinfo {author} {\bibfnamefont {R.~F.}\ \bibnamefont
  {Sabirianov}}, \bibinfo {author} {\bibfnamefont {S.~S.}\ \bibnamefont
  {Jaswal}},\ and\ \bibinfo {author} {\bibfnamefont {E.~Y.}\ \bibnamefont
  {Tsymbal}},\ }\bibfield  {title} {\bibinfo {title} {Giant electroresistance
  in ferroelectric tunnel junctions},\ }\href
  {https://doi.org/10.1103/PhysRevLett.94.246802} {\bibfield  {journal}
  {\bibinfo  {journal} {Phys. Rev. Lett.}\ }\textbf {\bibinfo {volume} {94}},\
  \bibinfo {pages} {246802} (\bibinfo {year} {2005})}\BibitemShut {NoStop}%
\bibitem [{\citenamefont {Wang}\ \emph {et~al.}(2020)\citenamefont {Wang},
  \citenamefont {Li}, \citenamefont {Hu}, \citenamefont {Fan}, \citenamefont
  {Fan}, \citenamefont {Li}, \citenamefont {Zhang}, \citenamefont {Liu},
  \citenamefont {Fan}, \citenamefont {Hou}, \citenamefont {Dang}, \citenamefont
  {Kou},\ and\ \citenamefont {Guo}}]{wang_direct_2020}%
  \BibitemOpen
  \bibfield  {author} {\bibinfo {author} {\bibfnamefont {Z.}~\bibnamefont
  {Wang}}, \bibinfo {author} {\bibfnamefont {H.}~\bibnamefont {Li}}, \bibinfo
  {author} {\bibfnamefont {H.}~\bibnamefont {Hu}}, \bibinfo {author}
  {\bibfnamefont {Y.}~\bibnamefont {Fan}}, \bibinfo {author} {\bibfnamefont
  {R.}~\bibnamefont {Fan}}, \bibinfo {author} {\bibfnamefont {B.}~\bibnamefont
  {Li}}, \bibinfo {author} {\bibfnamefont {J.}~\bibnamefont {Zhang}}, \bibinfo
  {author} {\bibfnamefont {H.}~\bibnamefont {Liu}}, \bibinfo {author}
  {\bibfnamefont {J.}~\bibnamefont {Fan}}, \bibinfo {author} {\bibfnamefont
  {H.}~\bibnamefont {Hou}}, \bibinfo {author} {\bibfnamefont {F.}~\bibnamefont
  {Dang}}, \bibinfo {author} {\bibfnamefont {Z.}~\bibnamefont {Kou}},\ and\
  \bibinfo {author} {\bibfnamefont {Z.}~\bibnamefont {Guo}},\ }\bibfield
  {title} {\bibinfo {title} {Direct observation of stable negative capacitance
  in {SrTiO3$@$BaTiO$_3$} heterostructure},\ }\href
  {https://doi.org/https://doi.org/10.1002/aelm.201901005} {\bibfield
  {journal} {\bibinfo  {journal} {Advanced Electronic Materials}\ }\textbf
  {\bibinfo {volume} {6}},\ \bibinfo {pages} {1901005} (\bibinfo {year}
  {2020})}\BibitemShut {NoStop}%
\bibitem [{\citenamefont {Yan}\ \emph {et~al.}(2011)\citenamefont {Yan},
  \citenamefont {Guo}, \citenamefont {Zhang},\ and\ \citenamefont
  {Liu}}]{yan_high-performance_2011}%
  \BibitemOpen
  \bibfield  {author} {\bibinfo {author} {\bibfnamefont {Z.}~\bibnamefont
  {Yan}}, \bibinfo {author} {\bibfnamefont {Y.}~\bibnamefont {Guo}}, \bibinfo
  {author} {\bibfnamefont {G.}~\bibnamefont {Zhang}},\ and\ \bibinfo {author}
  {\bibfnamefont {J.-M.}\ \bibnamefont {Liu}},\ }\bibfield  {title} {\bibinfo
  {title} {High-performance programmable memory devices based on co-doped
  {BaTiO$_3$}},\ }\href
  {https://doi.org/https://doi.org/10.1002/adma.201004306} {\bibfield
  {journal} {\bibinfo  {journal} {Advanced Materials}\ }\textbf {\bibinfo
  {volume} {23}},\ \bibinfo {pages} {1351} (\bibinfo {year}
  {2011})}\BibitemShut {NoStop}%
\bibitem [{\citenamefont {Rafiq}\ \emph {et~al.}(2022)\citenamefont {Rafiq},
  \citenamefont {Parihar}, \citenamefont {Chauhan},\ and\ \citenamefont
  {Sahay}}]{Musaib_FE}%
  \BibitemOpen
  \bibfield  {author} {\bibinfo {author} {\bibfnamefont {M.}~\bibnamefont
  {Rafiq}}, \bibinfo {author} {\bibfnamefont {S.~S.}\ \bibnamefont {Parihar}},
  \bibinfo {author} {\bibfnamefont {Y.~S.}\ \bibnamefont {Chauhan}},\ and\
  \bibinfo {author} {\bibfnamefont {S.}~\bibnamefont {Sahay}},\ }\bibfield
  {title} {\bibinfo {title} {Efficient implementation of max-pooling algorithm
  exploiting history-effect in ferroelectric-finfets},\ }\href
  {https://doi.org/10.1109/TED.2022.3207114} {\bibfield  {journal} {\bibinfo
  {journal} {IEEE Transactions on Electron Devices}\ }\textbf {\bibinfo
  {volume} {69}},\ \bibinfo {pages} {6446} (\bibinfo {year}
  {2022})}\BibitemShut {NoStop}%
\bibitem [{\citenamefont {Wang}\ \emph {et~al.}(2014)\citenamefont {Wang},
  \citenamefont {Liu}, \citenamefont {Man~Lau}, \citenamefont {Wang},
  \citenamefont {Yang}, \citenamefont {Zheng},\ and\ \citenamefont
  {Li}}]{wang_field-effect_2014}%
  \BibitemOpen
  \bibfield  {author} {\bibinfo {author} {\bibfnamefont {W.}~\bibnamefont
  {Wang}}, \bibinfo {author} {\bibfnamefont {F.}~\bibnamefont {Liu}}, \bibinfo
  {author} {\bibfnamefont {C.}~\bibnamefont {Man~Lau}}, \bibinfo {author}
  {\bibfnamefont {L.}~\bibnamefont {Wang}}, \bibinfo {author} {\bibfnamefont
  {G.}~\bibnamefont {Yang}}, \bibinfo {author} {\bibfnamefont {D.}~\bibnamefont
  {Zheng}},\ and\ \bibinfo {author} {\bibfnamefont {Z.}~\bibnamefont {Li}},\
  }\bibfield  {title} {\bibinfo {title} {Field-effect {BaTiO$_3$-Si} solar
  cells},\ }\href {https://doi.org/10.1063/1.4869556} {\bibfield  {journal}
  {\bibinfo  {journal} {Applied Physics Letters}\ }\textbf {\bibinfo {volume}
  {104}},\ \bibinfo {pages} {123901} (\bibinfo {year} {2014})}\BibitemShut
  {NoStop}%
\bibitem [{\citenamefont {Priydarshi}\ \emph {et~al.}(2022)\citenamefont
  {Priydarshi}, \citenamefont {Chauhan}, \citenamefont {Bhowmick},\ and\
  \citenamefont {Agarwal}}]{Achintya}%
  \BibitemOpen
  \bibfield  {author} {\bibinfo {author} {\bibfnamefont {A.}~\bibnamefont
  {Priydarshi}}, \bibinfo {author} {\bibfnamefont {Y.~S.}\ \bibnamefont
  {Chauhan}}, \bibinfo {author} {\bibfnamefont {S.}~\bibnamefont {Bhowmick}},\
  and\ \bibinfo {author} {\bibfnamefont {A.}~\bibnamefont {Agarwal}},\
  }\bibfield  {title} {\bibinfo {title} {{Strain-tunable in-plane
  ferroelectricity and lateral tunnel junction in monolayer group-IV
  monochalcogenides}},\ }\href {https://doi.org/10.1063/5.0072124} {\bibfield
  {journal} {\bibinfo  {journal} {Journal of Applied Physics}\ }\textbf
  {\bibinfo {volume} {131}},\ \bibinfo {pages} {034101} (\bibinfo {year}
  {2022})}\BibitemShut {NoStop}%
\bibitem [{\citenamefont {Rafiq}\ \emph {et~al.}(2023)\citenamefont {Rafiq},
  \citenamefont {Kaur}, \citenamefont {Gaidhane}, \citenamefont {Chauhan},\
  and\ \citenamefont {Sahay}}]{Musaib_FE1}%
  \BibitemOpen
  \bibfield  {author} {\bibinfo {author} {\bibfnamefont {M.}~\bibnamefont
  {Rafiq}}, \bibinfo {author} {\bibfnamefont {T.}~\bibnamefont {Kaur}},
  \bibinfo {author} {\bibfnamefont {A.}~\bibnamefont {Gaidhane}}, \bibinfo
  {author} {\bibfnamefont {Y.~S.}\ \bibnamefont {Chauhan}},\ and\ \bibinfo
  {author} {\bibfnamefont {S.}~\bibnamefont {Sahay}},\ }\bibfield  {title}
  {\bibinfo {title} {Ferroelectric fet-based time-mode multiply-accumulate
  accelerator: Design and analysis},\ }\href
  {https://doi.org/10.1109/TED.2023.3323261} {\bibfield  {journal} {\bibinfo
  {journal} {IEEE Transactions on Electron Devices}\ }\textbf {\bibinfo
  {volume} {70}},\ \bibinfo {pages} {6613} (\bibinfo {year}
  {2023})}\BibitemShut {NoStop}%
\bibitem [{\citenamefont {Hoffmann}\ \emph {et~al.}(2019)\citenamefont
  {Hoffmann}, \citenamefont {Fengler}, \citenamefont {Herzig}, \citenamefont
  {Mittmann}, \citenamefont {Max}, \citenamefont {Schroeder}, \citenamefont
  {Negrea}, \citenamefont {Lucian}, \citenamefont {Slesazeck},\ and\
  \citenamefont {Mikolajick}}]{hoffmann_unveiling_2019}%
  \BibitemOpen
  \bibfield  {author} {\bibinfo {author} {\bibfnamefont {M.}~\bibnamefont
  {Hoffmann}}, \bibinfo {author} {\bibfnamefont {F.~P.~G.}\ \bibnamefont
  {Fengler}}, \bibinfo {author} {\bibfnamefont {M.}~\bibnamefont {Herzig}},
  \bibinfo {author} {\bibfnamefont {T.}~\bibnamefont {Mittmann}}, \bibinfo
  {author} {\bibfnamefont {B.}~\bibnamefont {Max}}, \bibinfo {author}
  {\bibfnamefont {U.}~\bibnamefont {Schroeder}}, \bibinfo {author}
  {\bibfnamefont {R.}~\bibnamefont {Negrea}}, \bibinfo {author} {\bibfnamefont
  {P.}~\bibnamefont {Lucian}}, \bibinfo {author} {\bibfnamefont
  {S.}~\bibnamefont {Slesazeck}},\ and\ \bibinfo {author} {\bibfnamefont
  {T.}~\bibnamefont {Mikolajick}},\ }\bibfield  {title} {\bibinfo {title}
  {Unveiling the double-well energy landscape in a ferroelectric layer},\
  }\href {https://doi.org/10.1038/s41586-018-0854-z} {\bibfield  {journal}
  {\bibinfo  {journal} {Nature}\ }\textbf {\bibinfo {volume} {565}},\ \bibinfo
  {pages} {464} (\bibinfo {year} {2019})}\BibitemShut {NoStop}%
\bibitem [{\citenamefont {Dawber}\ \emph {et~al.}(2005)\citenamefont {Dawber},
  \citenamefont {Rabe},\ and\ \citenamefont {Scott}}]{dep_1}%
  \BibitemOpen
  \bibfield  {author} {\bibinfo {author} {\bibfnamefont {M.}~\bibnamefont
  {Dawber}}, \bibinfo {author} {\bibfnamefont {K.~M.}\ \bibnamefont {Rabe}},\
  and\ \bibinfo {author} {\bibfnamefont {J.~F.}\ \bibnamefont {Scott}},\
  }\bibfield  {title} {\bibinfo {title} {Physics of thin-film ferroelectric
  oxides},\ }\href {https://doi.org/10.1103/RevModPhys.77.1083} {\bibfield
  {journal} {\bibinfo  {journal} {Rev. Mod. Phys.}\ }\textbf {\bibinfo {volume}
  {77}},\ \bibinfo {pages} {1083} (\bibinfo {year} {2005})}\BibitemShut
  {NoStop}%
\bibitem [{\citenamefont {Zhong}\ \emph {et~al.}(1994)\citenamefont {Zhong},
  \citenamefont {King-Smith},\ and\ \citenamefont {Vanderbilt}}]{dep_2}%
  \BibitemOpen
  \bibfield  {author} {\bibinfo {author} {\bibfnamefont {W.}~\bibnamefont
  {Zhong}}, \bibinfo {author} {\bibfnamefont {R.~D.}\ \bibnamefont
  {King-Smith}},\ and\ \bibinfo {author} {\bibfnamefont {D.}~\bibnamefont
  {Vanderbilt}},\ }\bibfield  {title} {\bibinfo {title} {Giant lo-to splittings
  in perovskite ferroelectrics},\ }\href
  {https://doi.org/10.1103/PhysRevLett.72.3618} {\bibfield  {journal} {\bibinfo
   {journal} {Phys. Rev. Lett.}\ }\textbf {\bibinfo {volume} {72}},\ \bibinfo
  {pages} {3618} (\bibinfo {year} {1994})}\BibitemShut {NoStop}%
\bibitem [{\citenamefont {Fong}\ \emph {et~al.}(2004)\citenamefont {Fong},
  \citenamefont {Stephenson}, \citenamefont {Streiffer}, \citenamefont
  {Eastman}, \citenamefont {Auciello}, \citenamefont {Fuoss},\ and\
  \citenamefont {Thompson}}]{dep_3}%
  \BibitemOpen
  \bibfield  {author} {\bibinfo {author} {\bibfnamefont {D.~D.}\ \bibnamefont
  {Fong}}, \bibinfo {author} {\bibfnamefont {G.~B.}\ \bibnamefont
  {Stephenson}}, \bibinfo {author} {\bibfnamefont {S.~K.}\ \bibnamefont
  {Streiffer}}, \bibinfo {author} {\bibfnamefont {J.~A.}\ \bibnamefont
  {Eastman}}, \bibinfo {author} {\bibfnamefont {O.}~\bibnamefont {Auciello}},
  \bibinfo {author} {\bibfnamefont {P.~H.}\ \bibnamefont {Fuoss}},\ and\
  \bibinfo {author} {\bibfnamefont {C.}~\bibnamefont {Thompson}},\ }\bibfield
  {title} {\bibinfo {title} {Ferroelectricity in ultrathin perovskite films},\
  }\href {https://doi.org/10.1126/science.1098252} {\bibfield  {journal}
  {\bibinfo  {journal} {Science}\ }\textbf {\bibinfo {volume} {304}},\ \bibinfo
  {pages} {1650} (\bibinfo {year} {2004})}\BibitemShut {NoStop}%
\bibitem [{\citenamefont {Junquera}\ and\ \citenamefont
  {Ghosez}(2003)}]{dep_4}%
  \BibitemOpen
  \bibfield  {author} {\bibinfo {author} {\bibfnamefont {J.}~\bibnamefont
  {Junquera}}\ and\ \bibinfo {author} {\bibfnamefont {P.}~\bibnamefont
  {Ghosez}},\ }\bibfield  {title} {\bibinfo {title} {Critical thickness for
  ferroelectricity in perovskite ultrathin films},\ }\href
  {https://doi.org/10.1038/nature01501} {\bibfield  {journal} {\bibinfo
  {journal} {Nature}\ }\textbf {\bibinfo {volume} {422}},\ \bibinfo {pages}
  {506} (\bibinfo {year} {2003})}\BibitemShut {NoStop}%
\bibitem [{\citenamefont {Ahn}\ \emph {et~al.}(2004)\citenamefont {Ahn},
  \citenamefont {Rabe},\ and\ \citenamefont {Triscone}}]{dep_5}%
  \BibitemOpen
  \bibfield  {author} {\bibinfo {author} {\bibfnamefont {C.~H.}\ \bibnamefont
  {Ahn}}, \bibinfo {author} {\bibfnamefont {K.~M.}\ \bibnamefont {Rabe}},\ and\
  \bibinfo {author} {\bibfnamefont {J.-M.}\ \bibnamefont {Triscone}},\
  }\bibfield  {title} {\bibinfo {title} {Ferroelectricity at the nanoscale:
  Local polarization in oxide thin films and heterostructures},\ }\href
  {https://doi.org/10.1126/science.1092508} {\bibfield  {journal} {\bibinfo
  {journal} {Science}\ }\textbf {\bibinfo {volume} {303}},\ \bibinfo {pages}
  {488} (\bibinfo {year} {2004})}\BibitemShut {NoStop}%
\bibitem [{\citenamefont {Chhowalla}\ \emph {et~al.}(2016)\citenamefont
  {Chhowalla}, \citenamefont {Jena},\ and\ \citenamefont
  {Zhang}}]{Chhowalla2016}%
  \BibitemOpen
  \bibfield  {author} {\bibinfo {author} {\bibfnamefont {M.}~\bibnamefont
  {Chhowalla}}, \bibinfo {author} {\bibfnamefont {D.}~\bibnamefont {Jena}},\
  and\ \bibinfo {author} {\bibfnamefont {H.}~\bibnamefont {Zhang}},\ }\bibfield
   {title} {\bibinfo {title} {Two-dimensional semiconductors for transistors},\
  }\href {https://doi.org/10.1038/natrevmats.2016.52} {\bibfield  {journal}
  {\bibinfo  {journal} {Nature Reviews Materials}\ }\textbf {\bibinfo {volume}
  {1}},\ \bibinfo {pages} {16052} (\bibinfo {year} {2016})}\BibitemShut
  {NoStop}%
\bibitem [{\citenamefont {Das}\ \emph {et~al.}(2021)\citenamefont {Das},
  \citenamefont {Sebastian}, \citenamefont {Pop}, \citenamefont {McClellan},
  \citenamefont {Franklin}, \citenamefont {Grasser}, \citenamefont {Knobloch},
  \citenamefont {Illarionov}, \citenamefont {Penumatcha}, \citenamefont
  {Appenzeller}, \citenamefont {Chen}, \citenamefont {Zhu}, \citenamefont
  {Asselberghs}, \citenamefont {Li}, \citenamefont {Avci}, \citenamefont
  {Bhat}, \citenamefont {Anthopoulos},\ and\ \citenamefont
  {Singh}}]{Das2021_2D}%
  \BibitemOpen
  \bibfield  {author} {\bibinfo {author} {\bibfnamefont {S.}~\bibnamefont
  {Das}}, \bibinfo {author} {\bibfnamefont {A.}~\bibnamefont {Sebastian}},
  \bibinfo {author} {\bibfnamefont {E.}~\bibnamefont {Pop}}, \bibinfo {author}
  {\bibfnamefont {C.~J.}\ \bibnamefont {McClellan}}, \bibinfo {author}
  {\bibfnamefont {A.~D.}\ \bibnamefont {Franklin}}, \bibinfo {author}
  {\bibfnamefont {T.}~\bibnamefont {Grasser}}, \bibinfo {author} {\bibfnamefont
  {T.}~\bibnamefont {Knobloch}}, \bibinfo {author} {\bibfnamefont
  {Y.}~\bibnamefont {Illarionov}}, \bibinfo {author} {\bibfnamefont {A.~V.}\
  \bibnamefont {Penumatcha}}, \bibinfo {author} {\bibfnamefont
  {J.}~\bibnamefont {Appenzeller}}, \bibinfo {author} {\bibfnamefont
  {Z.}~\bibnamefont {Chen}}, \bibinfo {author} {\bibfnamefont {W.}~\bibnamefont
  {Zhu}}, \bibinfo {author} {\bibfnamefont {I.}~\bibnamefont {Asselberghs}},
  \bibinfo {author} {\bibfnamefont {L.-J.}\ \bibnamefont {Li}}, \bibinfo
  {author} {\bibfnamefont {U.~E.}\ \bibnamefont {Avci}}, \bibinfo {author}
  {\bibfnamefont {N.}~\bibnamefont {Bhat}}, \bibinfo {author} {\bibfnamefont
  {T.~D.}\ \bibnamefont {Anthopoulos}},\ and\ \bibinfo {author} {\bibfnamefont
  {R.}~\bibnamefont {Singh}},\ }\bibfield  {title} {\bibinfo {title}
  {Transistors based on two-dimensional materials for future integrated
  circuits},\ }\href {https://doi.org/10.1038/s41928-021-00670-1} {\bibfield
  {journal} {\bibinfo  {journal} {Nature Electronics}\ }\textbf {\bibinfo
  {volume} {4}},\ \bibinfo {pages} {786} (\bibinfo {year} {2021})}\BibitemShut
  {NoStop}%
\bibitem [{\citenamefont {Nandan}\ \emph {et~al.}(2023)\citenamefont {Nandan},
  \citenamefont {Naseer},\ and\ \citenamefont {Chauhan}}]{Nandan2023}%
  \BibitemOpen
  \bibfield  {author} {\bibinfo {author} {\bibfnamefont {K.}~\bibnamefont
  {Nandan}}, \bibinfo {author} {\bibfnamefont {A.}~\bibnamefont {Naseer}},\
  and\ \bibinfo {author} {\bibfnamefont {Y.~S.}\ \bibnamefont {Chauhan}},\
  }\bibfield  {title} {\bibinfo {title} {Field-effect transistors based on
  two-dimensional materials (invited)},\ }\href
  {https://doi.org/10.1007/s41403-022-00379-3} {\bibfield  {journal} {\bibinfo
  {journal} {Transactions of the Indian National Academy of Engineering}\
  }\textbf {\bibinfo {volume} {8}},\ \bibinfo {pages} {1} (\bibinfo {year}
  {2023})}\BibitemShut {NoStop}%
\bibitem [{\citenamefont {Novoselov}(2011)}]{Nov}%
  \BibitemOpen
  \bibfield  {author} {\bibinfo {author} {\bibfnamefont {K.~S.}\ \bibnamefont
  {Novoselov}},\ }\bibfield  {title} {\bibinfo {title} {Nobel lecture:
  Graphene: Materials in the flatland},\ }\href
  {https://doi.org/10.1103/RevModPhys.83.837} {\bibfield  {journal} {\bibinfo
  {journal} {Rev. Mod. Phys.}\ }\textbf {\bibinfo {volume} {83}},\ \bibinfo
  {pages} {837} (\bibinfo {year} {2011})}\BibitemShut {NoStop}%
\bibitem [{\citenamefont {Novoselov}\ \emph {et~al.}(2005)\citenamefont
  {Novoselov}, \citenamefont {Jiang}, \citenamefont {Schedin}, \citenamefont
  {Booth}, \citenamefont {Khotkevich}, \citenamefont {Morozov},\ and\
  \citenamefont {Geim}}]{Nov1}%
  \BibitemOpen
  \bibfield  {author} {\bibinfo {author} {\bibfnamefont {K.~S.}\ \bibnamefont
  {Novoselov}}, \bibinfo {author} {\bibfnamefont {D.}~\bibnamefont {Jiang}},
  \bibinfo {author} {\bibfnamefont {F.}~\bibnamefont {Schedin}}, \bibinfo
  {author} {\bibfnamefont {T.~J.}\ \bibnamefont {Booth}}, \bibinfo {author}
  {\bibfnamefont {V.~V.}\ \bibnamefont {Khotkevich}}, \bibinfo {author}
  {\bibfnamefont {S.~V.}\ \bibnamefont {Morozov}},\ and\ \bibinfo {author}
  {\bibfnamefont {A.~K.}\ \bibnamefont {Geim}},\ }\bibfield  {title} {\bibinfo
  {title} {Two-dimensional atomic crystals},\ }\href
  {https://doi.org/10.1073/pnas.0502848102} {\bibfield  {journal} {\bibinfo
  {journal} {Proceedings of the National Academy of Sciences}\ }\textbf
  {\bibinfo {volume} {102}},\ \bibinfo {pages} {10451} (\bibinfo {year}
  {2005})}\BibitemShut {NoStop}%
\bibitem [{\citenamefont {Naseer}\ \emph
  {et~al.}(2023{\natexlab{a}})\citenamefont {Naseer}, \citenamefont {Nandan},
  \citenamefont {Agarwal}, \citenamefont {Bhowmick},\ and\ \citenamefont
  {Chauhan}}]{Ateeb_1}%
  \BibitemOpen
  \bibfield  {author} {\bibinfo {author} {\bibfnamefont {A.}~\bibnamefont
  {Naseer}}, \bibinfo {author} {\bibfnamefont {K.}~\bibnamefont {Nandan}},
  \bibinfo {author} {\bibfnamefont {A.}~\bibnamefont {Agarwal}}, \bibinfo
  {author} {\bibfnamefont {S.}~\bibnamefont {Bhowmick}},\ and\ \bibinfo
  {author} {\bibfnamefont {Y.~S.}\ \bibnamefont {Chauhan}},\ }\bibfield
  {title} {\bibinfo {title} {Di-metal chalcogenides: A new family of promising
  {2-D} semiconductors for high-performance transistors},\ }\href
  {https://doi.org/10.1109/TED.2023.3261831} {\bibfield  {journal} {\bibinfo
  {journal} {IEEE Transactions on Electron Devices}\ }\textbf {\bibinfo
  {volume} {70}},\ \bibinfo {pages} {2445} (\bibinfo {year}
  {2023}{\natexlab{a}})}\BibitemShut {NoStop}%
\bibitem [{\citenamefont {Yadav}\ \emph {et~al.}(2017)\citenamefont {Yadav},
  \citenamefont {Agarwal},\ and\ \citenamefont {Chauhan}}]{Yadav_TMDs}%
  \BibitemOpen
  \bibfield  {author} {\bibinfo {author} {\bibfnamefont {C.}~\bibnamefont
  {Yadav}}, \bibinfo {author} {\bibfnamefont {A.}~\bibnamefont {Agarwal}},\
  and\ \bibinfo {author} {\bibfnamefont {Y.~S.}\ \bibnamefont {Chauhan}},\
  }\bibfield  {title} {\bibinfo {title} {Compact modeling of transition metal
  dichalcogenide based thin body transistors and circuit validation},\ }\href
  {https://doi.org/10.1109/TED.2016.2643698} {\bibfield  {journal} {\bibinfo
  {journal} {IEEE Transactions on Electron Devices}\ }\textbf {\bibinfo
  {volume} {64}},\ \bibinfo {pages} {1261} (\bibinfo {year}
  {2017})}\BibitemShut {NoStop}%
\bibitem [{\citenamefont {Yoon}\ \emph {et~al.}(2011)\citenamefont {Yoon},
  \citenamefont {Ganapathi},\ and\ \citenamefont {Salahuddin}}]{Yoon}%
  \BibitemOpen
  \bibfield  {author} {\bibinfo {author} {\bibfnamefont {Y.}~\bibnamefont
  {Yoon}}, \bibinfo {author} {\bibfnamefont {K.}~\bibnamefont {Ganapathi}},\
  and\ \bibinfo {author} {\bibfnamefont {S.}~\bibnamefont {Salahuddin}},\
  }\bibfield  {title} {\bibinfo {title} {How good can monolayer {M}o{S$_{2}$}
  transistors be?},\ }\href {https://doi.org/10.1021/nl2018178} {\bibfield
  {journal} {\bibinfo  {journal} {Nano Letters}\ }\textbf {\bibinfo {volume}
  {11}},\ \bibinfo {pages} {3768} (\bibinfo {year} {2011})}\BibitemShut
  {NoStop}%
\bibitem [{\citenamefont {Kou}\ \emph {et~al.}(2015)\citenamefont {Kou},
  \citenamefont {Chen},\ and\ \citenamefont {Smith}}]{Kou}%
  \BibitemOpen
  \bibfield  {author} {\bibinfo {author} {\bibfnamefont {L.}~\bibnamefont
  {Kou}}, \bibinfo {author} {\bibfnamefont {C.}~\bibnamefont {Chen}},\ and\
  \bibinfo {author} {\bibfnamefont {S.~C.}\ \bibnamefont {Smith}},\ }\bibfield
  {title} {\bibinfo {title} {Phosphorene: Fabrication, properties, and
  applications},\ }\href {https://doi.org/10.1021/acs.jpclett.5b01094}
  {\bibfield  {journal} {\bibinfo  {journal} {The Journal of Physical Chemistry
  Letters}\ }\textbf {\bibinfo {volume} {6}},\ \bibinfo {pages} {2794}
  (\bibinfo {year} {2015})}\BibitemShut {NoStop}%
\bibitem [{\citenamefont {Rodin}\ \emph {et~al.}(2014)\citenamefont {Rodin},
  \citenamefont {Carvalho},\ and\ \citenamefont {Castro~Neto}}]{Rodin}%
  \BibitemOpen
  \bibfield  {author} {\bibinfo {author} {\bibfnamefont {A.~S.}\ \bibnamefont
  {Rodin}}, \bibinfo {author} {\bibfnamefont {A.}~\bibnamefont {Carvalho}},\
  and\ \bibinfo {author} {\bibfnamefont {A.~H.}\ \bibnamefont {Castro~Neto}},\
  }\bibfield  {title} {\bibinfo {title} {Strain-induced gap modification in
  black phosphorus},\ }\href {https://doi.org/10.1103/PhysRevLett.112.176801}
  {\bibfield  {journal} {\bibinfo  {journal} {Phys. Rev. Lett.}\ }\textbf
  {\bibinfo {volume} {112}},\ \bibinfo {pages} {176801} (\bibinfo {year}
  {2014})}\BibitemShut {NoStop}%
\bibitem [{\citenamefont {Naseer}\ \emph
  {et~al.}(2023{\natexlab{b}})\citenamefont {Naseer}, \citenamefont {Nandan},
  \citenamefont {Agarwal}, \citenamefont {Bhowmick},\ and\ \citenamefont
  {Chauhan}}]{Ateeb2}%
  \BibitemOpen
  \bibfield  {author} {\bibinfo {author} {\bibfnamefont {A.}~\bibnamefont
  {Naseer}}, \bibinfo {author} {\bibfnamefont {K.}~\bibnamefont {Nandan}},
  \bibinfo {author} {\bibfnamefont {A.}~\bibnamefont {Agarwal}}, \bibinfo
  {author} {\bibfnamefont {S.}~\bibnamefont {Bhowmick}},\ and\ \bibinfo
  {author} {\bibfnamefont {Y.~S.}\ \bibnamefont {Chauhan}},\ }\bibfield
  {title} {\bibinfo {title} {Performance evaluation of monolayer {ZrS}$_{3}$
  transistors for next-generation computing},\ }\href
  {https://doi.org/10.1109/TED.2023.3304273} {\bibfield  {journal} {\bibinfo
  {journal} {IEEE Transactions on Electron Devices}\ }\textbf {\bibinfo
  {volume} {70}},\ \bibinfo {pages} {5435} (\bibinfo {year}
  {2023}{\natexlab{b}})}\BibitemShut {NoStop}%
\bibitem [{\citenamefont {Belianinov}\ \emph {et~al.}(2015)\citenamefont
  {Belianinov}, \citenamefont {He}, \citenamefont {Dziaugys}, \citenamefont
  {Maksymovych}, \citenamefont {Eliseev}, \citenamefont {Borisevich},
  \citenamefont {Morozovska}, \citenamefont {Banys}, \citenamefont
  {Vysochanskii},\ and\ \citenamefont {Kalinin}}]{belianinov_cuinp2s6_2015}%
  \BibitemOpen
  \bibfield  {author} {\bibinfo {author} {\bibfnamefont {A.}~\bibnamefont
  {Belianinov}}, \bibinfo {author} {\bibfnamefont {Q.}~\bibnamefont {He}},
  \bibinfo {author} {\bibfnamefont {A.}~\bibnamefont {Dziaugys}}, \bibinfo
  {author} {\bibfnamefont {P.}~\bibnamefont {Maksymovych}}, \bibinfo {author}
  {\bibfnamefont {E.}~\bibnamefont {Eliseev}}, \bibinfo {author} {\bibfnamefont
  {A.}~\bibnamefont {Borisevich}}, \bibinfo {author} {\bibfnamefont
  {A.}~\bibnamefont {Morozovska}}, \bibinfo {author} {\bibfnamefont
  {J.}~\bibnamefont {Banys}}, \bibinfo {author} {\bibfnamefont
  {Y.}~\bibnamefont {Vysochanskii}},\ and\ \bibinfo {author} {\bibfnamefont
  {S.~V.}\ \bibnamefont {Kalinin}},\ }\bibfield  {title} {\bibinfo {title}
  {{CuInP$_2$S$_6$} room temperature layered ferroelectric},\ }\href
  {https://doi.org/10.1021/acs.nanolett.5b00491} {\bibfield  {journal}
  {\bibinfo  {journal} {Nano Letters}\ }\textbf {\bibinfo {volume} {15}},\
  \bibinfo {pages} {3808} (\bibinfo {year} {2015})}\BibitemShut {NoStop}%
\bibitem [{\citenamefont {Liu}\ \emph {et~al.}(2016)\citenamefont {Liu},
  \citenamefont {You}, \citenamefont {Seyler}, \citenamefont {Li},
  \citenamefont {Yu}, \citenamefont {Lin}, \citenamefont {Wang}, \citenamefont
  {Zhou}, \citenamefont {Wang}, \citenamefont {He}, \citenamefont {Pantelides},
  \citenamefont {Zhou}, \citenamefont {Sharma}, \citenamefont {Xu},
  \citenamefont {Ajayan}, \citenamefont {Wang},\ and\ \citenamefont
  {Liu}}]{liu_room-temperature_2016}%
  \BibitemOpen
  \bibfield  {author} {\bibinfo {author} {\bibfnamefont {F.}~\bibnamefont
  {Liu}}, \bibinfo {author} {\bibfnamefont {L.}~\bibnamefont {You}}, \bibinfo
  {author} {\bibfnamefont {K.~L.}\ \bibnamefont {Seyler}}, \bibinfo {author}
  {\bibfnamefont {X.}~\bibnamefont {Li}}, \bibinfo {author} {\bibfnamefont
  {P.}~\bibnamefont {Yu}}, \bibinfo {author} {\bibfnamefont {J.}~\bibnamefont
  {Lin}}, \bibinfo {author} {\bibfnamefont {X.}~\bibnamefont {Wang}}, \bibinfo
  {author} {\bibfnamefont {J.}~\bibnamefont {Zhou}}, \bibinfo {author}
  {\bibfnamefont {H.}~\bibnamefont {Wang}}, \bibinfo {author} {\bibfnamefont
  {H.}~\bibnamefont {He}}, \bibinfo {author} {\bibfnamefont {S.~T.}\
  \bibnamefont {Pantelides}}, \bibinfo {author} {\bibfnamefont
  {W.}~\bibnamefont {Zhou}}, \bibinfo {author} {\bibfnamefont {P.}~\bibnamefont
  {Sharma}}, \bibinfo {author} {\bibfnamefont {X.}~\bibnamefont {Xu}}, \bibinfo
  {author} {\bibfnamefont {P.~M.}\ \bibnamefont {Ajayan}}, \bibinfo {author}
  {\bibfnamefont {J.}~\bibnamefont {Wang}},\ and\ \bibinfo {author}
  {\bibfnamefont {Z.}~\bibnamefont {Liu}},\ }\bibfield  {title} {\bibinfo
  {title} {Room-temperature ferroelectricity in {CuInP$_2$S$_6$} ultrathin
  flakes},\ }\href {https://doi.org/10.1038/ncomms12357} {\bibfield  {journal}
  {\bibinfo  {journal} {Nature Communications}\ }\textbf {\bibinfo {volume}
  {7}},\ \bibinfo {pages} {12357} (\bibinfo {year} {2016})}\BibitemShut
  {NoStop}%
\bibitem [{\citenamefont {Chang}\ \emph {et~al.}(2016)\citenamefont {Chang},
  \citenamefont {Liu}, \citenamefont {Lin}, \citenamefont {Wang}, \citenamefont
  {Zhao}, \citenamefont {Zhang}, \citenamefont {Jin}, \citenamefont {Zhong},
  \citenamefont {Hu}, \citenamefont {Duan}, \citenamefont {Zhang},
  \citenamefont {Fu}, \citenamefont {Xue}, \citenamefont {Chen},\ and\
  \citenamefont {Ji}}]{chang_discovery_2016}%
  \BibitemOpen
  \bibfield  {author} {\bibinfo {author} {\bibfnamefont {K.}~\bibnamefont
  {Chang}}, \bibinfo {author} {\bibfnamefont {J.}~\bibnamefont {Liu}}, \bibinfo
  {author} {\bibfnamefont {H.}~\bibnamefont {Lin}}, \bibinfo {author}
  {\bibfnamefont {N.}~\bibnamefont {Wang}}, \bibinfo {author} {\bibfnamefont
  {K.}~\bibnamefont {Zhao}}, \bibinfo {author} {\bibfnamefont {A.}~\bibnamefont
  {Zhang}}, \bibinfo {author} {\bibfnamefont {F.}~\bibnamefont {Jin}}, \bibinfo
  {author} {\bibfnamefont {Y.}~\bibnamefont {Zhong}}, \bibinfo {author}
  {\bibfnamefont {X.}~\bibnamefont {Hu}}, \bibinfo {author} {\bibfnamefont
  {W.}~\bibnamefont {Duan}}, \bibinfo {author} {\bibfnamefont {Q.}~\bibnamefont
  {Zhang}}, \bibinfo {author} {\bibfnamefont {L.}~\bibnamefont {Fu}}, \bibinfo
  {author} {\bibfnamefont {Q.-K.}\ \bibnamefont {Xue}}, \bibinfo {author}
  {\bibfnamefont {X.}~\bibnamefont {Chen}},\ and\ \bibinfo {author}
  {\bibfnamefont {S.-H.}\ \bibnamefont {Ji}},\ }\bibfield  {title} {\bibinfo
  {title} {Discovery of robust in-plane ferroelectricity in atomic-thick
  {SnTe}},\ }\href {https://doi.org/10.1126/science.aad8609} {\bibfield
  {journal} {\bibinfo  {journal} {Science}\ }\textbf {\bibinfo {volume}
  {353}},\ \bibinfo {pages} {274} (\bibinfo {year} {2016})}\BibitemShut
  {NoStop}%
\bibitem [{\citenamefont {Cui}\ \emph {et~al.}(2018)\citenamefont {Cui},
  \citenamefont {Hu}, \citenamefont {Yan}, \citenamefont {Addiego},
  \citenamefont {Gao}, \citenamefont {Wang}, \citenamefont {Wang},
  \citenamefont {Li}, \citenamefont {Cheng}, \citenamefont {Li}, \citenamefont
  {Zhang}, \citenamefont {Alshareef}, \citenamefont {Wu}, \citenamefont {Zhu},
  \citenamefont {Pan},\ and\ \citenamefont {Li}}]{cui_intercorrelated_2018}%
  \BibitemOpen
  \bibfield  {author} {\bibinfo {author} {\bibfnamefont {C.}~\bibnamefont
  {Cui}}, \bibinfo {author} {\bibfnamefont {W.-J.}\ \bibnamefont {Hu}},
  \bibinfo {author} {\bibfnamefont {X.}~\bibnamefont {Yan}}, \bibinfo {author}
  {\bibfnamefont {C.}~\bibnamefont {Addiego}}, \bibinfo {author} {\bibfnamefont
  {W.}~\bibnamefont {Gao}}, \bibinfo {author} {\bibfnamefont {Y.}~\bibnamefont
  {Wang}}, \bibinfo {author} {\bibfnamefont {Z.}~\bibnamefont {Wang}}, \bibinfo
  {author} {\bibfnamefont {L.}~\bibnamefont {Li}}, \bibinfo {author}
  {\bibfnamefont {Y.}~\bibnamefont {Cheng}}, \bibinfo {author} {\bibfnamefont
  {P.}~\bibnamefont {Li}}, \bibinfo {author} {\bibfnamefont {X.}~\bibnamefont
  {Zhang}}, \bibinfo {author} {\bibfnamefont {H.~N.}\ \bibnamefont
  {Alshareef}}, \bibinfo {author} {\bibfnamefont {T.}~\bibnamefont {Wu}},
  \bibinfo {author} {\bibfnamefont {W.}~\bibnamefont {Zhu}}, \bibinfo {author}
  {\bibfnamefont {X.}~\bibnamefont {Pan}},\ and\ \bibinfo {author}
  {\bibfnamefont {L.-J.}\ \bibnamefont {Li}},\ }\bibfield  {title} {\bibinfo
  {title} {Intercorrelated in-plane and out-of-plane ferroelectricity in
  ultrathin two-dimensional layered semiconductor {In$_2$Se$_3$}},\ }\href
  {https://doi.org/10.1021/acs.nanolett.7b04852} {\bibfield  {journal}
  {\bibinfo  {journal} {Nano Letters}\ }\textbf {\bibinfo {volume} {18}},\
  \bibinfo {pages} {1253} (\bibinfo {year} {2018})}\BibitemShut {NoStop}%
\bibitem [{\citenamefont {Xiao}\ \emph
  {et~al.}(2018{\natexlab{a}})\citenamefont {Xiao}, \citenamefont {Zhu},
  \citenamefont {Wang}, \citenamefont {Feng}, \citenamefont {Hu}, \citenamefont
  {Dasgupta}, \citenamefont {Han}, \citenamefont {Wang}, \citenamefont
  {Muller}, \citenamefont {Martin}, \citenamefont {Hu},\ and\ \citenamefont
  {Zhang}}]{xiao_intrinsic_2018}%
  \BibitemOpen
  \bibfield  {author} {\bibinfo {author} {\bibfnamefont {J.}~\bibnamefont
  {Xiao}}, \bibinfo {author} {\bibfnamefont {H.}~\bibnamefont {Zhu}}, \bibinfo
  {author} {\bibfnamefont {Y.}~\bibnamefont {Wang}}, \bibinfo {author}
  {\bibfnamefont {W.}~\bibnamefont {Feng}}, \bibinfo {author} {\bibfnamefont
  {Y.}~\bibnamefont {Hu}}, \bibinfo {author} {\bibfnamefont {A.}~\bibnamefont
  {Dasgupta}}, \bibinfo {author} {\bibfnamefont {Y.}~\bibnamefont {Han}},
  \bibinfo {author} {\bibfnamefont {Y.}~\bibnamefont {Wang}}, \bibinfo {author}
  {\bibfnamefont {D.~A.}\ \bibnamefont {Muller}}, \bibinfo {author}
  {\bibfnamefont {L.~W.}\ \bibnamefont {Martin}}, \bibinfo {author}
  {\bibfnamefont {P.}~\bibnamefont {Hu}},\ and\ \bibinfo {author}
  {\bibfnamefont {X.}~\bibnamefont {Zhang}},\ }\bibfield  {title} {\bibinfo
  {title} {Intrinsic two-dimensional ferroelectricity with dipole locking},\
  }\href {https://doi.org/10.1103/PhysRevLett.120.227601} {\bibfield  {journal}
  {\bibinfo  {journal} {Phys. Rev. Lett.}\ }\textbf {\bibinfo {volume} {120}},\
  \bibinfo {pages} {227601} (\bibinfo {year} {2018}{\natexlab{a}})}\BibitemShut
  {NoStop}%
\bibitem [{\citenamefont {Wan}\ \emph {et~al.}(2018)\citenamefont {Wan},
  \citenamefont {Li}, \citenamefont {Li}, \citenamefont {Mao}, \citenamefont
  {Zhu},\ and\ \citenamefont {Zeng}}]{wan_room-temperature_2018}%
  \BibitemOpen
  \bibfield  {author} {\bibinfo {author} {\bibfnamefont {S.}~\bibnamefont
  {Wan}}, \bibinfo {author} {\bibfnamefont {Y.}~\bibnamefont {Li}}, \bibinfo
  {author} {\bibfnamefont {W.}~\bibnamefont {Li}}, \bibinfo {author}
  {\bibfnamefont {X.}~\bibnamefont {Mao}}, \bibinfo {author} {\bibfnamefont
  {W.}~\bibnamefont {Zhu}},\ and\ \bibinfo {author} {\bibfnamefont
  {H.}~\bibnamefont {Zeng}},\ }\bibfield  {title} {\bibinfo {title}
  {{Room-temperature ferroelectricity and a switchable diode effect in
  two-dimensional $\alpha-$In$_2$Se$_3$ thin layers}},\ }\href
  {https://doi.org/10.1039/C8NR04422H} {\bibfield  {journal} {\bibinfo
  {journal} {Nanoscale}\ }\textbf {\bibinfo {volume} {10}},\ \bibinfo {pages}
  {14885} (\bibinfo {year} {2018})}\BibitemShut {NoStop}%
\bibitem [{\citenamefont {Poh}\ \emph {et~al.}(2018)\citenamefont {Poh},
  \citenamefont {Tan}, \citenamefont {Wang}, \citenamefont {Song},
  \citenamefont {Abidi}, \citenamefont {Zhao}, \citenamefont {Dan},
  \citenamefont {Chen}, \citenamefont {Luo}, \citenamefont {Pennycook},
  \citenamefont {Castro~Neto},\ and\ \citenamefont
  {Loh}}]{poh_molecular-beam_2018}%
  \BibitemOpen
  \bibfield  {author} {\bibinfo {author} {\bibfnamefont {S.~M.}\ \bibnamefont
  {Poh}}, \bibinfo {author} {\bibfnamefont {S.~J.~R.}\ \bibnamefont {Tan}},
  \bibinfo {author} {\bibfnamefont {H.}~\bibnamefont {Wang}}, \bibinfo {author}
  {\bibfnamefont {P.}~\bibnamefont {Song}}, \bibinfo {author} {\bibfnamefont
  {I.~H.}\ \bibnamefont {Abidi}}, \bibinfo {author} {\bibfnamefont
  {X.}~\bibnamefont {Zhao}}, \bibinfo {author} {\bibfnamefont {J.}~\bibnamefont
  {Dan}}, \bibinfo {author} {\bibfnamefont {J.}~\bibnamefont {Chen}}, \bibinfo
  {author} {\bibfnamefont {Z.}~\bibnamefont {Luo}}, \bibinfo {author}
  {\bibfnamefont {S.~J.}\ \bibnamefont {Pennycook}}, \bibinfo {author}
  {\bibfnamefont {A.~H.}\ \bibnamefont {Castro~Neto}},\ and\ \bibinfo {author}
  {\bibfnamefont {K.~P.}\ \bibnamefont {Loh}},\ }\bibfield  {title} {\bibinfo
  {title} {Molecular-beam epitaxy of two-dimensional {In$_2$Se$_3$} and its
  giant electroresistance switching in ferroresistive memory junction},\ }\href
  {https://doi.org/10.1021/acs.nanolett.8b02688} {\bibfield  {journal}
  {\bibinfo  {journal} {Nano Letters}\ }\textbf {\bibinfo {volume} {18}},\
  \bibinfo {pages} {6340} (\bibinfo {year} {2018})}\BibitemShut {NoStop}%
\bibitem [{\citenamefont {Fei}\ \emph {et~al.}(2018)\citenamefont {Fei},
  \citenamefont {Zhao}, \citenamefont {Palomaki}, \citenamefont {Sun},
  \citenamefont {Miller}, \citenamefont {Zhao}, \citenamefont {Yan},
  \citenamefont {Xu},\ and\ \citenamefont {Cobden}}]{fei_ferroelectric_2018}%
  \BibitemOpen
  \bibfield  {author} {\bibinfo {author} {\bibfnamefont {Z.}~\bibnamefont
  {Fei}}, \bibinfo {author} {\bibfnamefont {W.}~\bibnamefont {Zhao}}, \bibinfo
  {author} {\bibfnamefont {T.~A.}\ \bibnamefont {Palomaki}}, \bibinfo {author}
  {\bibfnamefont {B.}~\bibnamefont {Sun}}, \bibinfo {author} {\bibfnamefont
  {M.~K.}\ \bibnamefont {Miller}}, \bibinfo {author} {\bibfnamefont
  {Z.}~\bibnamefont {Zhao}}, \bibinfo {author} {\bibfnamefont {J.}~\bibnamefont
  {Yan}}, \bibinfo {author} {\bibfnamefont {X.}~\bibnamefont {Xu}},\ and\
  \bibinfo {author} {\bibfnamefont {D.~H.}\ \bibnamefont {Cobden}},\ }\bibfield
   {title} {\bibinfo {title} {Ferroelectric switching of a two-dimensional
  metal},\ }\href {https://doi.org/10.1038/s41586-018-0336-3} {\bibfield
  {journal} {\bibinfo  {journal} {Nature}\ }\textbf {\bibinfo {volume} {560}},\
  \bibinfo {pages} {336} (\bibinfo {year} {2018})}\BibitemShut {NoStop}%
\bibitem [{\citenamefont {Zheng}\ \emph {et~al.}(2018)\citenamefont {Zheng},
  \citenamefont {Yu}, \citenamefont {Zhu}, \citenamefont {Collins},
  \citenamefont {Kim}, \citenamefont {Lou}, \citenamefont {Xu}, \citenamefont
  {Li}, \citenamefont {Wei}, \citenamefont {Zhang}, \citenamefont {Edmonds},
  \citenamefont {Li}, \citenamefont {Seidel}, \citenamefont {Zhu},
  \citenamefont {Liu}, \citenamefont {Tang},\ and\ \citenamefont
  {Fuhrer}}]{zheng_room_2018}%
  \BibitemOpen
  \bibfield  {author} {\bibinfo {author} {\bibfnamefont {C.}~\bibnamefont
  {Zheng}}, \bibinfo {author} {\bibfnamefont {L.}~\bibnamefont {Yu}}, \bibinfo
  {author} {\bibfnamefont {L.}~\bibnamefont {Zhu}}, \bibinfo {author}
  {\bibfnamefont {J.~L.}\ \bibnamefont {Collins}}, \bibinfo {author}
  {\bibfnamefont {D.}~\bibnamefont {Kim}}, \bibinfo {author} {\bibfnamefont
  {Y.}~\bibnamefont {Lou}}, \bibinfo {author} {\bibfnamefont {C.}~\bibnamefont
  {Xu}}, \bibinfo {author} {\bibfnamefont {M.}~\bibnamefont {Li}}, \bibinfo
  {author} {\bibfnamefont {Z.}~\bibnamefont {Wei}}, \bibinfo {author}
  {\bibfnamefont {Y.}~\bibnamefont {Zhang}}, \bibinfo {author} {\bibfnamefont
  {M.~T.}\ \bibnamefont {Edmonds}}, \bibinfo {author} {\bibfnamefont
  {S.}~\bibnamefont {Li}}, \bibinfo {author} {\bibfnamefont {J.}~\bibnamefont
  {Seidel}}, \bibinfo {author} {\bibfnamefont {Y.}~\bibnamefont {Zhu}},
  \bibinfo {author} {\bibfnamefont {J.~Z.}\ \bibnamefont {Liu}}, \bibinfo
  {author} {\bibfnamefont {W.-X.}\ \bibnamefont {Tang}},\ and\ \bibinfo
  {author} {\bibfnamefont {M.~S.}\ \bibnamefont {Fuhrer}},\ }\bibfield  {title}
  {\bibinfo {title} {Room temperature in-plane ferroelectricity in van der
  waals {In$_2$Se$_3$}},\ }\href {https://doi.org/10.1126/sciadv.aar7720}
  {\bibfield  {journal} {\bibinfo  {journal} {Science Advances}\ }\textbf
  {\bibinfo {volume} {4}},\ \bibinfo {pages} {eaar7720} (\bibinfo {year}
  {2018})}\BibitemShut {NoStop}%
\bibitem [{\citenamefont {Xiao}\ \emph
  {et~al.}(2020{\natexlab{a}})\citenamefont {Xiao}, \citenamefont {Shao},
  \citenamefont {Zhang},\ and\ \citenamefont
  {Jiang}}]{PhysRevApplied.13.044014}%
  \BibitemOpen
  \bibfield  {author} {\bibinfo {author} {\bibfnamefont {R.-C.}\ \bibnamefont
  {Xiao}}, \bibinfo {author} {\bibfnamefont {D.-F.}\ \bibnamefont {Shao}},
  \bibinfo {author} {\bibfnamefont {Z.-Q.}\ \bibnamefont {Zhang}},\ and\
  \bibinfo {author} {\bibfnamefont {H.}~\bibnamefont {Jiang}},\ }\bibfield
  {title} {\bibinfo {title} {Two-dimensional metals for piezoelectriclike
  devices based on berry-curvature dipole},\ }\href
  {https://doi.org/10.1103/PhysRevApplied.13.044014} {\bibfield  {journal}
  {\bibinfo  {journal} {Phys. Rev. Appl.}\ }\textbf {\bibinfo {volume} {13}},\
  \bibinfo {pages} {044014} (\bibinfo {year} {2020}{\natexlab{a}})}\BibitemShut
  {NoStop}%
\bibitem [{\citenamefont {Mahajan}\ and\ \citenamefont
  {Bhowmick}(2021)}]{PhysRevB.103.075436}%
  \BibitemOpen
  \bibfield  {author} {\bibinfo {author} {\bibfnamefont {A.}~\bibnamefont
  {Mahajan}}\ and\ \bibinfo {author} {\bibfnamefont {S.}~\bibnamefont
  {Bhowmick}},\ }\bibfield  {title} {\bibinfo {title} {Decoupled strain
  response of ferroic properties in a multiferroic ${\mathrm{vocl}}_{2}$
  monolayer},\ }\href {https://doi.org/10.1103/PhysRevB.103.075436} {\bibfield
  {journal} {\bibinfo  {journal} {Phys. Rev. B}\ }\textbf {\bibinfo {volume}
  {103}},\ \bibinfo {pages} {075436} (\bibinfo {year} {2021})}\BibitemShut
  {NoStop}%
\bibitem [{\citenamefont {Kang}\ \emph
  {et~al.}(2019{\natexlab{a}})\citenamefont {Kang}, \citenamefont {Jiang},
  \citenamefont {Cao}, \citenamefont {Hao}, \citenamefont {Zheng},
  \citenamefont {Zhang},\ and\ \citenamefont {Zeng}}]{kang_realizing_2019}%
  \BibitemOpen
  \bibfield  {author} {\bibinfo {author} {\bibfnamefont {L.}~\bibnamefont
  {Kang}}, \bibinfo {author} {\bibfnamefont {P.}~\bibnamefont {Jiang}},
  \bibinfo {author} {\bibfnamefont {N.}~\bibnamefont {Cao}}, \bibinfo {author}
  {\bibfnamefont {H.}~\bibnamefont {Hao}}, \bibinfo {author} {\bibfnamefont
  {X.}~\bibnamefont {Zheng}}, \bibinfo {author} {\bibfnamefont
  {L.}~\bibnamefont {Zhang}},\ and\ \bibinfo {author} {\bibfnamefont
  {Z.}~\bibnamefont {Zeng}},\ }\bibfield  {title} {\bibinfo {title} {Realizing
  giant tunneling electroresistance in two-dimensional {graphene/BiP}
  ferroelectric tunnel junction},\ }\href {https://doi.org/10.1039/C9NR01656B}
  {\bibfield  {journal} {\bibinfo  {journal} {Nanoscale}\ }\textbf {\bibinfo
  {volume} {11}},\ \bibinfo {pages} {16837} (\bibinfo {year}
  {2019}{\natexlab{a}})}\BibitemShut {NoStop}%
\bibitem [{\citenamefont {Shen}\ \emph {et~al.}(2019)\citenamefont {Shen},
  \citenamefont {Fang}, \citenamefont {Tian},\ and\ \citenamefont
  {Duan}}]{shen_two-dimensional_2019}%
  \BibitemOpen
  \bibfield  {author} {\bibinfo {author} {\bibfnamefont {X.-W.}\ \bibnamefont
  {Shen}}, \bibinfo {author} {\bibfnamefont {Y.-W.}\ \bibnamefont {Fang}},
  \bibinfo {author} {\bibfnamefont {B.-B.}\ \bibnamefont {Tian}},\ and\
  \bibinfo {author} {\bibfnamefont {C.-G.}\ \bibnamefont {Duan}},\ }\bibfield
  {title} {\bibinfo {title} {Two-dimensional ferroelectric tunnel junction: The
  case of monolayer {In:SnSe/SnSe/Sb:SnSe} homostructure},\ }\href
  {https://doi.org/10.1021/acsaelm.9b00146} {\bibfield  {journal} {\bibinfo
  {journal} {ACS Applied Electronic Materials}\ }\textbf {\bibinfo {volume}
  {1}},\ \bibinfo {pages} {1133} (\bibinfo {year} {2019})}\BibitemShut
  {NoStop}%
\bibitem [{\citenamefont {Sodemann}\ and\ \citenamefont
  {Fu}(2015)}]{PhysRevLett.115.216806}%
  \BibitemOpen
  \bibfield  {author} {\bibinfo {author} {\bibfnamefont {I.}~\bibnamefont
  {Sodemann}}\ and\ \bibinfo {author} {\bibfnamefont {L.}~\bibnamefont {Fu}},\
  }\bibfield  {title} {\bibinfo {title} {Quantum nonlinear hall effect induced
  by berry curvature dipole in time-reversal invariant materials},\ }\href
  {https://doi.org/10.1103/PhysRevLett.115.216806} {\bibfield  {journal}
  {\bibinfo  {journal} {Phys. Rev. Lett.}\ }\textbf {\bibinfo {volume} {115}},\
  \bibinfo {pages} {216806} (\bibinfo {year} {2015})}\BibitemShut {NoStop}%
\bibitem [{\citenamefont {Low}\ \emph {et~al.}(2015)\citenamefont {Low},
  \citenamefont {Jiang},\ and\ \citenamefont {Guinea}}]{PhysRevB.92.235447}%
  \BibitemOpen
  \bibfield  {author} {\bibinfo {author} {\bibfnamefont {T.}~\bibnamefont
  {Low}}, \bibinfo {author} {\bibfnamefont {Y.}~\bibnamefont {Jiang}},\ and\
  \bibinfo {author} {\bibfnamefont {F.}~\bibnamefont {Guinea}},\ }\bibfield
  {title} {\bibinfo {title} {Topological currents in black phosphorus with
  broken inversion symmetry},\ }\href
  {https://doi.org/10.1103/PhysRevB.92.235447} {\bibfield  {journal} {\bibinfo
  {journal} {Phys. Rev. B}\ }\textbf {\bibinfo {volume} {92}},\ \bibinfo
  {pages} {235447} (\bibinfo {year} {2015})}\BibitemShut {NoStop}%
\bibitem [{\citenamefont {Sinha}\ \emph {et~al.}(2022)\citenamefont {Sinha},
  \citenamefont {Adak}, \citenamefont {Chakraborty}, \citenamefont {Das},
  \citenamefont {Debnath}, \citenamefont {Sangani}, \citenamefont {Watanabe},
  \citenamefont {Taniguchi}, \citenamefont {Waghmare}, \citenamefont
  {Agarwal},\ and\ \citenamefont {Deshmukh}}]{Subhajit_NPhy_23}%
  \BibitemOpen
  \bibfield  {author} {\bibinfo {author} {\bibfnamefont {S.}~\bibnamefont
  {Sinha}}, \bibinfo {author} {\bibfnamefont {P.~C.}\ \bibnamefont {Adak}},
  \bibinfo {author} {\bibfnamefont {A.}~\bibnamefont {Chakraborty}}, \bibinfo
  {author} {\bibfnamefont {K.}~\bibnamefont {Das}}, \bibinfo {author}
  {\bibfnamefont {K.}~\bibnamefont {Debnath}}, \bibinfo {author} {\bibfnamefont
  {L.~D.~V.}\ \bibnamefont {Sangani}}, \bibinfo {author} {\bibfnamefont
  {K.}~\bibnamefont {Watanabe}}, \bibinfo {author} {\bibfnamefont
  {T.}~\bibnamefont {Taniguchi}}, \bibinfo {author} {\bibfnamefont {U.~V.}\
  \bibnamefont {Waghmare}}, \bibinfo {author} {\bibfnamefont {A.}~\bibnamefont
  {Agarwal}},\ and\ \bibinfo {author} {\bibfnamefont {M.~M.}\ \bibnamefont
  {Deshmukh}},\ }\bibfield  {title} {\bibinfo {title} {Berry curvature dipole
  senses topological transition in a moir{\'e}superlattice},\ }\href
  {https://doi.org/10.1038/s41567-022-01606-y} {\bibfield  {journal} {\bibinfo
  {journal} {Nature Physics}\ }\textbf {\bibinfo {volume} {18}},\ \bibinfo
  {pages} {765} (\bibinfo {year} {2022})}\BibitemShut {NoStop}%
\bibitem [{\citenamefont {Chakraborty}\ \emph {et~al.}(2022)\citenamefont
  {Chakraborty}, \citenamefont {Das}, \citenamefont {Sinha}, \citenamefont
  {Adak}, \citenamefont {Deshmukh},\ and\ \citenamefont
  {Agarwal}}]{Chakraborty_2022}%
  \BibitemOpen
  \bibfield  {author} {\bibinfo {author} {\bibfnamefont {A.}~\bibnamefont
  {Chakraborty}}, \bibinfo {author} {\bibfnamefont {K.}~\bibnamefont {Das}},
  \bibinfo {author} {\bibfnamefont {S.}~\bibnamefont {Sinha}}, \bibinfo
  {author} {\bibfnamefont {P.~C.}\ \bibnamefont {Adak}}, \bibinfo {author}
  {\bibfnamefont {M.~M.}\ \bibnamefont {Deshmukh}},\ and\ \bibinfo {author}
  {\bibfnamefont {A.}~\bibnamefont {Agarwal}},\ }\bibfield  {title} {\bibinfo
  {title} {Nonlinear anomalous hall effects probe topological phase-transitions
  in twisted double bilayer graphene},\ }\href
  {https://doi.org/10.1088/2053-1583/ac8b93} {\bibfield  {journal} {\bibinfo
  {journal} {2D Materials}\ }\textbf {\bibinfo {volume} {9}},\ \bibinfo {pages}
  {045020} (\bibinfo {year} {2022})}\BibitemShut {NoStop}%
\bibitem [{\citenamefont {Thouless}\ \emph {et~al.}(1982)\citenamefont
  {Thouless}, \citenamefont {Kohmoto}, \citenamefont {Nightingale},\ and\
  \citenamefont {den Nijs}}]{PhysRevLett.49.405}%
  \BibitemOpen
  \bibfield  {author} {\bibinfo {author} {\bibfnamefont {D.~J.}\ \bibnamefont
  {Thouless}}, \bibinfo {author} {\bibfnamefont {M.}~\bibnamefont {Kohmoto}},
  \bibinfo {author} {\bibfnamefont {M.~P.}\ \bibnamefont {Nightingale}},\ and\
  \bibinfo {author} {\bibfnamefont {M.}~\bibnamefont {den Nijs}},\ }\bibfield
  {title} {\bibinfo {title} {Quantized hall conductance in a two-dimensional
  periodic potential},\ }\href {https://doi.org/10.1103/PhysRevLett.49.405}
  {\bibfield  {journal} {\bibinfo  {journal} {Phys. Rev. Lett.}\ }\textbf
  {\bibinfo {volume} {49}},\ \bibinfo {pages} {405} (\bibinfo {year}
  {1982})}\BibitemShut {NoStop}%
\bibitem [{\citenamefont {Xiao}\ \emph {et~al.}(2010)\citenamefont {Xiao},
  \citenamefont {Chang},\ and\ \citenamefont {Niu}}]{RevModPhys.82.1959}%
  \BibitemOpen
  \bibfield  {author} {\bibinfo {author} {\bibfnamefont {D.}~\bibnamefont
  {Xiao}}, \bibinfo {author} {\bibfnamefont {M.-C.}\ \bibnamefont {Chang}},\
  and\ \bibinfo {author} {\bibfnamefont {Q.}~\bibnamefont {Niu}},\ }\bibfield
  {title} {\bibinfo {title} {Berry phase effects on electronic properties},\
  }\href {https://doi.org/10.1103/RevModPhys.82.1959} {\bibfield  {journal}
  {\bibinfo  {journal} {Rev. Mod. Phys.}\ }\textbf {\bibinfo {volume} {82}},\
  \bibinfo {pages} {1959} (\bibinfo {year} {2010})}\BibitemShut {NoStop}%
\bibitem [{\citenamefont {Yu}\ \emph {et~al.}(2010)\citenamefont {Yu},
  \citenamefont {Zhang}, \citenamefont {Zhang}, \citenamefont {Zhang},
  \citenamefont {Dai},\ and\ \citenamefont {Fang}}]{science.1187485}%
  \BibitemOpen
  \bibfield  {author} {\bibinfo {author} {\bibfnamefont {R.}~\bibnamefont
  {Yu}}, \bibinfo {author} {\bibfnamefont {W.}~\bibnamefont {Zhang}}, \bibinfo
  {author} {\bibfnamefont {H.-J.}\ \bibnamefont {Zhang}}, \bibinfo {author}
  {\bibfnamefont {S.-C.}\ \bibnamefont {Zhang}}, \bibinfo {author}
  {\bibfnamefont {X.}~\bibnamefont {Dai}},\ and\ \bibinfo {author}
  {\bibfnamefont {Z.}~\bibnamefont {Fang}},\ }\bibfield  {title} {\bibinfo
  {title} {Quantized anomalous hall effect in magnetic topological
  insulators},\ }\href {https://doi.org/10.1126/science.1187485} {\bibfield
  {journal} {\bibinfo  {journal} {Science}\ }\textbf {\bibinfo {volume}
  {329}},\ \bibinfo {pages} {61} (\bibinfo {year} {2010})}\BibitemShut
  {NoStop}%
\bibitem [{\citenamefont {Fang}\ \emph {et~al.}(2003)\citenamefont {Fang},
  \citenamefont {Nagaosa}, \citenamefont {Takahashi}, \citenamefont {Asamitsu},
  \citenamefont {Mathieu}, \citenamefont {Ogasawara}, \citenamefont {Yamada},
  \citenamefont {Kawasaki}, \citenamefont {Tokura},\ and\ \citenamefont
  {Terakura}}]{science.1089408}%
  \BibitemOpen
  \bibfield  {author} {\bibinfo {author} {\bibfnamefont {Z.}~\bibnamefont
  {Fang}}, \bibinfo {author} {\bibfnamefont {N.}~\bibnamefont {Nagaosa}},
  \bibinfo {author} {\bibfnamefont {K.~S.}\ \bibnamefont {Takahashi}}, \bibinfo
  {author} {\bibfnamefont {A.}~\bibnamefont {Asamitsu}}, \bibinfo {author}
  {\bibfnamefont {R.}~\bibnamefont {Mathieu}}, \bibinfo {author} {\bibfnamefont
  {T.}~\bibnamefont {Ogasawara}}, \bibinfo {author} {\bibfnamefont
  {H.}~\bibnamefont {Yamada}}, \bibinfo {author} {\bibfnamefont
  {M.}~\bibnamefont {Kawasaki}}, \bibinfo {author} {\bibfnamefont
  {Y.}~\bibnamefont {Tokura}},\ and\ \bibinfo {author} {\bibfnamefont
  {K.}~\bibnamefont {Terakura}},\ }\bibfield  {title} {\bibinfo {title} {The
  anomalous hall effect and magnetic monopoles in momentum space},\ }\href
  {https://doi.org/10.1126/science.1089408} {\bibfield  {journal} {\bibinfo
  {journal} {Science}\ }\textbf {\bibinfo {volume} {302}},\ \bibinfo {pages}
  {92} (\bibinfo {year} {2003})}\BibitemShut {NoStop}%
\bibitem [{\citenamefont {Wang}\ and\ \citenamefont {Qian}(2019)}]{Wang_2019}%
  \BibitemOpen
  \bibfield  {author} {\bibinfo {author} {\bibfnamefont {H.}~\bibnamefont
  {Wang}}\ and\ \bibinfo {author} {\bibfnamefont {X.}~\bibnamefont {Qian}},\
  }\bibfield  {title} {\bibinfo {title} {Ferroelectric nonlinear anomalous hall
  effect in few-layer wte2},\ }\href
  {https://doi.org/10.1038/s41524-019-0257-1} {\bibfield  {journal} {\bibinfo
  {journal} {npj Computational Materials}\ }\textbf {\bibinfo {volume} {5}},\
  \bibinfo {pages} {119} (\bibinfo {year} {2019})}\BibitemShut {NoStop}%
\bibitem [{\citenamefont {Xiao}\ \emph
  {et~al.}(2020{\natexlab{b}})\citenamefont {Xiao}, \citenamefont {Shao},
  \citenamefont {Huang},\ and\ \citenamefont {Jiang}}]{PhysRevB.102.024109}%
  \BibitemOpen
  \bibfield  {author} {\bibinfo {author} {\bibfnamefont {R.-C.}\ \bibnamefont
  {Xiao}}, \bibinfo {author} {\bibfnamefont {D.-F.}\ \bibnamefont {Shao}},
  \bibinfo {author} {\bibfnamefont {W.}~\bibnamefont {Huang}},\ and\ \bibinfo
  {author} {\bibfnamefont {H.}~\bibnamefont {Jiang}},\ }\bibfield  {title}
  {\bibinfo {title} {Electrical detection of ferroelectriclike metals through
  the nonlinear hall effect},\ }\href
  {https://doi.org/10.1103/PhysRevB.102.024109} {\bibfield  {journal} {\bibinfo
   {journal} {Phys. Rev. B}\ }\textbf {\bibinfo {volume} {102}},\ \bibinfo
  {pages} {024109} (\bibinfo {year} {2020}{\natexlab{b}})}\BibitemShut
  {NoStop}%
\bibitem [{\citenamefont {Kim}\ \emph {et~al.}(2019)\citenamefont {Kim},
  \citenamefont {Kim}, \citenamefont {Shin}, \citenamefont {Lee}, \citenamefont
  {Sinova}, \citenamefont {Park},\ and\ \citenamefont
  {Jin}}]{kim2019prediction}%
  \BibitemOpen
  \bibfield  {author} {\bibinfo {author} {\bibfnamefont {J.}~\bibnamefont
  {Kim}}, \bibinfo {author} {\bibfnamefont {K.-W.}\ \bibnamefont {Kim}},
  \bibinfo {author} {\bibfnamefont {D.}~\bibnamefont {Shin}}, \bibinfo {author}
  {\bibfnamefont {S.-H.}\ \bibnamefont {Lee}}, \bibinfo {author} {\bibfnamefont
  {J.}~\bibnamefont {Sinova}}, \bibinfo {author} {\bibfnamefont
  {N.}~\bibnamefont {Park}},\ and\ \bibinfo {author} {\bibfnamefont
  {H.}~\bibnamefont {Jin}},\ }\bibfield  {title} {\bibinfo {title} {Prediction
  of ferroelectricity-driven berry curvature enabling charge- and
  spin-controllable photocurrent in tin telluride monolayers},\ }\href
  {https://doi.org/10.1038/s41467-019-11964-6} {\bibfield  {journal} {\bibinfo
  {journal} {Nature Communications}\ }\textbf {\bibinfo {volume} {10}},\
  \bibinfo {pages} {3965} (\bibinfo {year} {2019})}\BibitemShut {NoStop}%
\bibitem [{\citenamefont {Isobe}\ \emph {et~al.}(2020)\citenamefont {Isobe},
  \citenamefont {Xu},\ and\ \citenamefont {Fu}}]{isobe2020high}%
  \BibitemOpen
  \bibfield  {author} {\bibinfo {author} {\bibfnamefont {H.}~\bibnamefont
  {Isobe}}, \bibinfo {author} {\bibfnamefont {S.-Y.}\ \bibnamefont {Xu}},\ and\
  \bibinfo {author} {\bibfnamefont {L.}~\bibnamefont {Fu}},\ }\bibfield
  {title} {\bibinfo {title} {High-frequency rectification via chiral bloch
  electrons},\ }\href {https://doi.org/10.1126/sciadv.aay2497} {\bibfield
  {journal} {\bibinfo  {journal} {Science Advances}\ }\textbf {\bibinfo
  {volume} {6}},\ \bibinfo {pages} {eaay2497} (\bibinfo {year}
  {2020})}\BibitemShut {NoStop}%
\bibitem [{\citenamefont {Ma}\ \emph {et~al.}(2019)\citenamefont {Ma},
  \citenamefont {Xu}, \citenamefont {Shen}, \citenamefont {MacNeill},
  \citenamefont {Fatemi}, \citenamefont {Chang}, \citenamefont {Mier~Valdivia},
  \citenamefont {Wu}, \citenamefont {Du}, \citenamefont {Hsu}, \citenamefont
  {Fang}, \citenamefont {Gibson}, \citenamefont {Watanabe}, \citenamefont
  {Taniguchi}, \citenamefont {Cava}, \citenamefont {Kaxiras}, \citenamefont
  {Lu}, \citenamefont {Lin}, \citenamefont {Fu}, \citenamefont {Gedik},\ and\
  \citenamefont {Jarillo-Herrero}}]{ma2019observation}%
  \BibitemOpen
  \bibfield  {author} {\bibinfo {author} {\bibfnamefont {Q.}~\bibnamefont
  {Ma}}, \bibinfo {author} {\bibfnamefont {S.-Y.}\ \bibnamefont {Xu}}, \bibinfo
  {author} {\bibfnamefont {H.}~\bibnamefont {Shen}}, \bibinfo {author}
  {\bibfnamefont {D.}~\bibnamefont {MacNeill}}, \bibinfo {author}
  {\bibfnamefont {V.}~\bibnamefont {Fatemi}}, \bibinfo {author} {\bibfnamefont
  {T.-R.}\ \bibnamefont {Chang}}, \bibinfo {author} {\bibfnamefont {A.~M.}\
  \bibnamefont {Mier~Valdivia}}, \bibinfo {author} {\bibfnamefont
  {S.}~\bibnamefont {Wu}}, \bibinfo {author} {\bibfnamefont {Z.}~\bibnamefont
  {Du}}, \bibinfo {author} {\bibfnamefont {C.-H.}\ \bibnamefont {Hsu}},
  \bibinfo {author} {\bibfnamefont {S.}~\bibnamefont {Fang}}, \bibinfo {author}
  {\bibfnamefont {Q.~D.}\ \bibnamefont {Gibson}}, \bibinfo {author}
  {\bibfnamefont {K.}~\bibnamefont {Watanabe}}, \bibinfo {author}
  {\bibfnamefont {T.}~\bibnamefont {Taniguchi}}, \bibinfo {author}
  {\bibfnamefont {R.~J.}\ \bibnamefont {Cava}}, \bibinfo {author}
  {\bibfnamefont {E.}~\bibnamefont {Kaxiras}}, \bibinfo {author} {\bibfnamefont
  {H.-Z.}\ \bibnamefont {Lu}}, \bibinfo {author} {\bibfnamefont
  {H.}~\bibnamefont {Lin}}, \bibinfo {author} {\bibfnamefont {L.}~\bibnamefont
  {Fu}}, \bibinfo {author} {\bibfnamefont {N.}~\bibnamefont {Gedik}},\ and\
  \bibinfo {author} {\bibfnamefont {P.}~\bibnamefont {Jarillo-Herrero}},\
  }\bibfield  {title} {\bibinfo {title} {Observation of the nonlinear hall
  effect under time-reversal-symmetric conditions},\ }\href
  {https://doi.org/10.1038/s41586-018-0807-6} {\bibfield  {journal} {\bibinfo
  {journal} {Nature}\ }\textbf {\bibinfo {volume} {565}},\ \bibinfo {pages}
  {337} (\bibinfo {year} {2019})}\BibitemShut {NoStop}%
\bibitem [{\citenamefont {Kang}\ \emph
  {et~al.}(2019{\natexlab{b}})\citenamefont {Kang}, \citenamefont {Li},
  \citenamefont {Sohn}, \citenamefont {Shan},\ and\ \citenamefont
  {Mak}}]{kang2019nonlinear}%
  \BibitemOpen
  \bibfield  {author} {\bibinfo {author} {\bibfnamefont {K.}~\bibnamefont
  {Kang}}, \bibinfo {author} {\bibfnamefont {T.}~\bibnamefont {Li}}, \bibinfo
  {author} {\bibfnamefont {E.}~\bibnamefont {Sohn}}, \bibinfo {author}
  {\bibfnamefont {J.}~\bibnamefont {Shan}},\ and\ \bibinfo {author}
  {\bibfnamefont {K.~F.}\ \bibnamefont {Mak}},\ }\bibfield  {title} {\bibinfo
  {title} {Nonlinear anomalous hall effect in few-layer {WTe}$_2$},\ }\href
  {https://doi.org/10.1038/s41563-019-0294-7} {\bibfield  {journal} {\bibinfo
  {journal} {Nature Materials}\ }\textbf {\bibinfo {volume} {18}},\ \bibinfo
  {pages} {324} (\bibinfo {year} {2019}{\natexlab{b}})}\BibitemShut {NoStop}%
\bibitem [{\citenamefont {Xu}\ \emph {et~al.}(2018)\citenamefont {Xu},
  \citenamefont {Ma}, \citenamefont {Shen}, \citenamefont {Fatemi},
  \citenamefont {Wu}, \citenamefont {Chang}, \citenamefont {Chang},
  \citenamefont {Valdivia}, \citenamefont {Chan}, \citenamefont {Gibson},
  \citenamefont {Zhou}, \citenamefont {Liu}, \citenamefont {Watanabe},
  \citenamefont {Taniguchi}, \citenamefont {Lin}, \citenamefont {Cava},
  \citenamefont {Fu}, \citenamefont {Gedik},\ and\ \citenamefont
  {Jarillo-Herrero}}]{xu2018electrically}%
  \BibitemOpen
  \bibfield  {author} {\bibinfo {author} {\bibfnamefont {S.-Y.}\ \bibnamefont
  {Xu}}, \bibinfo {author} {\bibfnamefont {Q.}~\bibnamefont {Ma}}, \bibinfo
  {author} {\bibfnamefont {H.}~\bibnamefont {Shen}}, \bibinfo {author}
  {\bibfnamefont {V.}~\bibnamefont {Fatemi}}, \bibinfo {author} {\bibfnamefont
  {S.}~\bibnamefont {Wu}}, \bibinfo {author} {\bibfnamefont {T.-R.}\
  \bibnamefont {Chang}}, \bibinfo {author} {\bibfnamefont {G.}~\bibnamefont
  {Chang}}, \bibinfo {author} {\bibfnamefont {A.~M.~M.}\ \bibnamefont
  {Valdivia}}, \bibinfo {author} {\bibfnamefont {C.-K.}\ \bibnamefont {Chan}},
  \bibinfo {author} {\bibfnamefont {Q.~D.}\ \bibnamefont {Gibson}}, \bibinfo
  {author} {\bibfnamefont {J.}~\bibnamefont {Zhou}}, \bibinfo {author}
  {\bibfnamefont {Z.}~\bibnamefont {Liu}}, \bibinfo {author} {\bibfnamefont
  {K.}~\bibnamefont {Watanabe}}, \bibinfo {author} {\bibfnamefont
  {T.}~\bibnamefont {Taniguchi}}, \bibinfo {author} {\bibfnamefont
  {H.}~\bibnamefont {Lin}}, \bibinfo {author} {\bibfnamefont {R.~J.}\
  \bibnamefont {Cava}}, \bibinfo {author} {\bibfnamefont {L.}~\bibnamefont
  {Fu}}, \bibinfo {author} {\bibfnamefont {N.}~\bibnamefont {Gedik}},\ and\
  \bibinfo {author} {\bibfnamefont {P.}~\bibnamefont {Jarillo-Herrero}},\
  }\bibfield  {title} {\bibinfo {title} {Electrically switchable berry
  curvature dipole in the monolayer topological insulator {WTe}$_2$},\ }\href
  {https://doi.org/10.1038/s41567-018-0189-6} {\bibfield  {journal} {\bibinfo
  {journal} {Nature Physics}\ }\textbf {\bibinfo {volume} {14}},\ \bibinfo
  {pages} {900} (\bibinfo {year} {2018})}\BibitemShut {NoStop}%
\bibitem [{\citenamefont {Giannozzi}\ \emph {et~al.}(2009)\citenamefont
  {Giannozzi}, \citenamefont {Baroni}, \citenamefont {Bonini}, \citenamefont
  {Calandra}, \citenamefont {Car}, \citenamefont {Cavazzoni}, \citenamefont
  {Ceresoli}, \citenamefont {Chiarotti}, \citenamefont {Cococcioni},
  \citenamefont {Dabo}, \citenamefont {Corso}, \citenamefont {de~Gironcoli},
  \citenamefont {Fabris}, \citenamefont {Fratesi}, \citenamefont {Gebauer},
  \citenamefont {Gerstmann}, \citenamefont {Gougoussis}, \citenamefont
  {Kokalj}, \citenamefont {Lazzeri}, \citenamefont {Martin-Samos},
  \citenamefont {Marzari}, \citenamefont {Mauri}, \citenamefont {Mazzarello},
  \citenamefont {Paolini}, \citenamefont {Pasquarello}, \citenamefont
  {Paulatto}, \citenamefont {Sbraccia}, \citenamefont {Scandolo}, \citenamefont
  {Sclauzero}, \citenamefont {Seitsonen}, \citenamefont {Smogunov},
  \citenamefont {Umari},\ and\ \citenamefont
  {Wentzcovitch}}]{giannozzi_quantum_2009}%
  \BibitemOpen
  \bibfield  {author} {\bibinfo {author} {\bibfnamefont {P.}~\bibnamefont
  {Giannozzi}}, \bibinfo {author} {\bibfnamefont {S.}~\bibnamefont {Baroni}},
  \bibinfo {author} {\bibfnamefont {N.}~\bibnamefont {Bonini}}, \bibinfo
  {author} {\bibfnamefont {M.}~\bibnamefont {Calandra}}, \bibinfo {author}
  {\bibfnamefont {R.}~\bibnamefont {Car}}, \bibinfo {author} {\bibfnamefont
  {C.}~\bibnamefont {Cavazzoni}}, \bibinfo {author} {\bibfnamefont
  {D.}~\bibnamefont {Ceresoli}}, \bibinfo {author} {\bibfnamefont {G.~L.}\
  \bibnamefont {Chiarotti}}, \bibinfo {author} {\bibfnamefont {M.}~\bibnamefont
  {Cococcioni}}, \bibinfo {author} {\bibfnamefont {I.}~\bibnamefont {Dabo}},
  \bibinfo {author} {\bibfnamefont {A.~D.}\ \bibnamefont {Corso}}, \bibinfo
  {author} {\bibfnamefont {S.}~\bibnamefont {de~Gironcoli}}, \bibinfo {author}
  {\bibfnamefont {S.}~\bibnamefont {Fabris}}, \bibinfo {author} {\bibfnamefont
  {G.}~\bibnamefont {Fratesi}}, \bibinfo {author} {\bibfnamefont
  {R.}~\bibnamefont {Gebauer}}, \bibinfo {author} {\bibfnamefont
  {U.}~\bibnamefont {Gerstmann}}, \bibinfo {author} {\bibfnamefont
  {C.}~\bibnamefont {Gougoussis}}, \bibinfo {author} {\bibfnamefont
  {A.}~\bibnamefont {Kokalj}}, \bibinfo {author} {\bibfnamefont
  {M.}~\bibnamefont {Lazzeri}}, \bibinfo {author} {\bibfnamefont
  {L.}~\bibnamefont {Martin-Samos}}, \bibinfo {author} {\bibfnamefont
  {N.}~\bibnamefont {Marzari}}, \bibinfo {author} {\bibfnamefont
  {F.}~\bibnamefont {Mauri}}, \bibinfo {author} {\bibfnamefont
  {R.}~\bibnamefont {Mazzarello}}, \bibinfo {author} {\bibfnamefont
  {S.}~\bibnamefont {Paolini}}, \bibinfo {author} {\bibfnamefont
  {A.}~\bibnamefont {Pasquarello}}, \bibinfo {author} {\bibfnamefont
  {L.}~\bibnamefont {Paulatto}}, \bibinfo {author} {\bibfnamefont
  {C.}~\bibnamefont {Sbraccia}}, \bibinfo {author} {\bibfnamefont
  {S.}~\bibnamefont {Scandolo}}, \bibinfo {author} {\bibfnamefont
  {G.}~\bibnamefont {Sclauzero}}, \bibinfo {author} {\bibfnamefont {A.~P.}\
  \bibnamefont {Seitsonen}}, \bibinfo {author} {\bibfnamefont {A.}~\bibnamefont
  {Smogunov}}, \bibinfo {author} {\bibfnamefont {P.}~\bibnamefont {Umari}},\
  and\ \bibinfo {author} {\bibfnamefont {R.~M.}\ \bibnamefont {Wentzcovitch}},\
  }\bibfield  {title} {\bibinfo {title} {Quantum espresso: a modular and
  open-source software project for quantum simulations of materials},\ }\href
  {https://doi.org/10.1088/0953-8984/21/39/395502} {\bibfield  {journal}
  {\bibinfo  {journal} {Journal of Physics: Condensed Matter}\ }\textbf
  {\bibinfo {volume} {21}},\ \bibinfo {pages} {395502} (\bibinfo {year}
  {2009})}\BibitemShut {NoStop}%
\bibitem [{\citenamefont {Giannozzi}\ \emph {et~al.}(2017)\citenamefont
  {Giannozzi}, \citenamefont {Andreussi}, \citenamefont {Brumme}, \citenamefont
  {Bunau}, \citenamefont {Nardelli}, \citenamefont {Calandra}, \citenamefont
  {Car}, \citenamefont {Cavazzoni}, \citenamefont {Ceresoli}, \citenamefont
  {Cococcioni}, \citenamefont {Colonna}, \citenamefont {Carnimeo},
  \citenamefont {Corso}, \citenamefont {de~Gironcoli}, \citenamefont {Delugas},
  \citenamefont {DiStasio}, \citenamefont {Ferretti}, \citenamefont {Floris},
  \citenamefont {Fratesi}, \citenamefont {Fugallo}, \citenamefont {Gebauer},
  \citenamefont {Gerstmann}, \citenamefont {Giustino}, \citenamefont {Gorni},
  \citenamefont {Jia}, \citenamefont {Kawamura}, \citenamefont {Ko},
  \citenamefont {Kokalj}, \citenamefont {Küçükbenli}, \citenamefont
  {Lazzeri}, \citenamefont {Marsili}, \citenamefont {Marzari}, \citenamefont
  {Mauri}, \citenamefont {Nguyen}, \citenamefont {Nguyen}, \citenamefont {de-la
  Roza}, \citenamefont {Paulatto}, \citenamefont {Poncé}, \citenamefont
  {Rocca}, \citenamefont {Sabatini}, \citenamefont {Santra}, \citenamefont
  {Schlipf}, \citenamefont {Seitsonen}, \citenamefont {Smogunov}, \citenamefont
  {Timrov}, \citenamefont {Thonhauser}, \citenamefont {Umari}, \citenamefont
  {Vast}, \citenamefont {Wu},\ and\ \citenamefont
  {Baroni}}]{giannozzi_advanced_2017}%
  \BibitemOpen
  \bibfield  {author} {\bibinfo {author} {\bibfnamefont {P.}~\bibnamefont
  {Giannozzi}}, \bibinfo {author} {\bibfnamefont {O.}~\bibnamefont
  {Andreussi}}, \bibinfo {author} {\bibfnamefont {T.}~\bibnamefont {Brumme}},
  \bibinfo {author} {\bibfnamefont {O.}~\bibnamefont {Bunau}}, \bibinfo
  {author} {\bibfnamefont {M.~B.}\ \bibnamefont {Nardelli}}, \bibinfo {author}
  {\bibfnamefont {M.}~\bibnamefont {Calandra}}, \bibinfo {author}
  {\bibfnamefont {R.}~\bibnamefont {Car}}, \bibinfo {author} {\bibfnamefont
  {C.}~\bibnamefont {Cavazzoni}}, \bibinfo {author} {\bibfnamefont
  {D.}~\bibnamefont {Ceresoli}}, \bibinfo {author} {\bibfnamefont
  {M.}~\bibnamefont {Cococcioni}}, \bibinfo {author} {\bibfnamefont
  {N.}~\bibnamefont {Colonna}}, \bibinfo {author} {\bibfnamefont
  {I.}~\bibnamefont {Carnimeo}}, \bibinfo {author} {\bibfnamefont {A.~D.}\
  \bibnamefont {Corso}}, \bibinfo {author} {\bibfnamefont {S.}~\bibnamefont
  {de~Gironcoli}}, \bibinfo {author} {\bibfnamefont {P.}~\bibnamefont
  {Delugas}}, \bibinfo {author} {\bibfnamefont {R.~A.}\ \bibnamefont
  {DiStasio}}, \bibinfo {author} {\bibfnamefont {A.}~\bibnamefont {Ferretti}},
  \bibinfo {author} {\bibfnamefont {A.}~\bibnamefont {Floris}}, \bibinfo
  {author} {\bibfnamefont {G.}~\bibnamefont {Fratesi}}, \bibinfo {author}
  {\bibfnamefont {G.}~\bibnamefont {Fugallo}}, \bibinfo {author} {\bibfnamefont
  {R.}~\bibnamefont {Gebauer}}, \bibinfo {author} {\bibfnamefont
  {U.}~\bibnamefont {Gerstmann}}, \bibinfo {author} {\bibfnamefont
  {F.}~\bibnamefont {Giustino}}, \bibinfo {author} {\bibfnamefont
  {T.}~\bibnamefont {Gorni}}, \bibinfo {author} {\bibfnamefont
  {J.}~\bibnamefont {Jia}}, \bibinfo {author} {\bibfnamefont {M.}~\bibnamefont
  {Kawamura}}, \bibinfo {author} {\bibfnamefont {H.-Y.}\ \bibnamefont {Ko}},
  \bibinfo {author} {\bibfnamefont {A.}~\bibnamefont {Kokalj}}, \bibinfo
  {author} {\bibfnamefont {E.}~\bibnamefont {Küçükbenli}}, \bibinfo {author}
  {\bibfnamefont {M.}~\bibnamefont {Lazzeri}}, \bibinfo {author} {\bibfnamefont
  {M.}~\bibnamefont {Marsili}}, \bibinfo {author} {\bibfnamefont
  {N.}~\bibnamefont {Marzari}}, \bibinfo {author} {\bibfnamefont
  {F.}~\bibnamefont {Mauri}}, \bibinfo {author} {\bibfnamefont {N.~L.}\
  \bibnamefont {Nguyen}}, \bibinfo {author} {\bibfnamefont {H.-V.}\
  \bibnamefont {Nguyen}}, \bibinfo {author} {\bibfnamefont {A.~O.}\
  \bibnamefont {de-la Roza}}, \bibinfo {author} {\bibfnamefont
  {L.}~\bibnamefont {Paulatto}}, \bibinfo {author} {\bibfnamefont
  {S.}~\bibnamefont {Poncé}}, \bibinfo {author} {\bibfnamefont
  {D.}~\bibnamefont {Rocca}}, \bibinfo {author} {\bibfnamefont
  {R.}~\bibnamefont {Sabatini}}, \bibinfo {author} {\bibfnamefont
  {B.}~\bibnamefont {Santra}}, \bibinfo {author} {\bibfnamefont
  {M.}~\bibnamefont {Schlipf}}, \bibinfo {author} {\bibfnamefont {A.~P.}\
  \bibnamefont {Seitsonen}}, \bibinfo {author} {\bibfnamefont {A.}~\bibnamefont
  {Smogunov}}, \bibinfo {author} {\bibfnamefont {I.}~\bibnamefont {Timrov}},
  \bibinfo {author} {\bibfnamefont {T.}~\bibnamefont {Thonhauser}}, \bibinfo
  {author} {\bibfnamefont {P.}~\bibnamefont {Umari}}, \bibinfo {author}
  {\bibfnamefont {N.}~\bibnamefont {Vast}}, \bibinfo {author} {\bibfnamefont
  {X.}~\bibnamefont {Wu}},\ and\ \bibinfo {author} {\bibfnamefont
  {S.}~\bibnamefont {Baroni}},\ }\bibfield  {title} {\bibinfo {title} {Advanced
  capabilities for materials modelling with quantum espresso},\ }\href
  {https://doi.org/10.1088/1361-648X/aa8f79} {\bibfield  {journal} {\bibinfo
  {journal} {Journal of Physics: Condensed Matter}\ }\textbf {\bibinfo {volume}
  {29}},\ \bibinfo {pages} {465901} (\bibinfo {year} {2017})}\BibitemShut
  {NoStop}%
\bibitem [{\citenamefont {Bl$\mathrm{\ddot {o}
  }$chl}(1994)}]{blochl_projector_1994}%
  \BibitemOpen
  \bibfield  {author} {\bibinfo {author} {\bibfnamefont {P.~E.}\ \bibnamefont
  {Bl$\mathrm{\ddot {o} }$chl}},\ }\bibfield  {title} {\bibinfo {title}
  {Projector augmented-wave method},\ }\href
  {https://doi.org/10.1103/PhysRevB.50.17953} {\bibfield  {journal} {\bibinfo
  {journal} {Phys. Rev. B}\ }\textbf {\bibinfo {volume} {50}},\ \bibinfo
  {pages} {17953} (\bibinfo {year} {1994})}\BibitemShut {NoStop}%
\bibitem [{\citenamefont {Momma}\ and\ \citenamefont
  {Izumi}(2011)}]{momma_vesta_2011}%
  \BibitemOpen
  \bibfield  {author} {\bibinfo {author} {\bibfnamefont {K.}~\bibnamefont
  {Momma}}\ and\ \bibinfo {author} {\bibfnamefont {F.}~\bibnamefont {Izumi}},\
  }\bibfield  {title} {\bibinfo {title} {{{\it VESTA3} for three-dimensional
  visualization of crystal, volumetric and morphology data}},\ }\href
  {https://doi.org/10.1107/S0021889811038970} {\bibfield  {journal} {\bibinfo
  {journal} {Journal of Applied Crystallography}\ }\textbf {\bibinfo {volume}
  {44}},\ \bibinfo {pages} {1272} (\bibinfo {year} {2011})}\BibitemShut
  {NoStop}%
\bibitem [{\citenamefont {Kokalj}(2003)}]{XCRYSDEN}%
  \BibitemOpen
  \bibfield  {author} {\bibinfo {author} {\bibfnamefont {A.}~\bibnamefont
  {Kokalj}},\ }\bibfield  {title} {\bibinfo {title} {Computer graphics and
  graphical user interfaces as tools in simulations of matter at the atomic
  scale},\ }\href
  {https://doi.org/https://doi.org/10.1016/S0927-0256(03)00104-6} {\bibfield
  {journal} {\bibinfo  {journal} {Computational Materials Science}\ }\textbf
  {\bibinfo {volume} {28}},\ \bibinfo {pages} {155} (\bibinfo {year} {2003})},\
  \bibinfo {note} {proceedings of the Symposium on Software Development for
  Process and Materials Design}\BibitemShut {NoStop}%
\bibitem [{\citenamefont {King-Smith}\ and\ \citenamefont
  {Vanderbilt}(1993)}]{Berry_1}%
  \BibitemOpen
  \bibfield  {author} {\bibinfo {author} {\bibfnamefont {R.~D.}\ \bibnamefont
  {King-Smith}}\ and\ \bibinfo {author} {\bibfnamefont {D.}~\bibnamefont
  {Vanderbilt}},\ }\bibfield  {title} {\bibinfo {title} {Theory of polarization
  of crystalline solids},\ }\href {https://doi.org/10.1103/PhysRevB.47.1651}
  {\bibfield  {journal} {\bibinfo  {journal} {Phys. Rev. B}\ }\textbf {\bibinfo
  {volume} {47}},\ \bibinfo {pages} {1651} (\bibinfo {year}
  {1993})}\BibitemShut {NoStop}%
\bibitem [{\citenamefont {Vanderbilt}\ and\ \citenamefont
  {King-Smith}(1993)}]{Berry_2}%
  \BibitemOpen
  \bibfield  {author} {\bibinfo {author} {\bibfnamefont {D.}~\bibnamefont
  {Vanderbilt}}\ and\ \bibinfo {author} {\bibfnamefont {R.~D.}\ \bibnamefont
  {King-Smith}},\ }\bibfield  {title} {\bibinfo {title} {Electric polarization
  as a bulk quantity and its relation to surface charge},\ }\href
  {https://doi.org/10.1103/PhysRevB.48.4442} {\bibfield  {journal} {\bibinfo
  {journal} {Phys. Rev. B}\ }\textbf {\bibinfo {volume} {48}},\ \bibinfo
  {pages} {4442} (\bibinfo {year} {1993})}\BibitemShut {NoStop}%
\bibitem [{\citenamefont {Vanderbilt}(2000)}]{Berry_3}%
  \BibitemOpen
  \bibfield  {author} {\bibinfo {author} {\bibfnamefont {D.}~\bibnamefont
  {Vanderbilt}},\ }\bibfield  {title} {\bibinfo {title} {Berry-phase theory of
  proper piezoelectric response},\ }\href
  {https://doi.org/https://doi.org/10.1016/S0022-3697(99)00273-5} {\bibfield
  {journal} {\bibinfo  {journal} {Journal of Physics and Chemistry of Solids}\
  }\textbf {\bibinfo {volume} {61}},\ \bibinfo {pages} {147} (\bibinfo {year}
  {2000})}\BibitemShut {NoStop}%
\bibitem [{\citenamefont {Di~Sante}\ \emph {et~al.}(2015)\citenamefont
  {Di~Sante}, \citenamefont {Stroppa}, \citenamefont {Barone}, \citenamefont
  {Whangbo},\ and\ \citenamefont {Picozzi}}]{di_sante_emergence_2015}%
  \BibitemOpen
  \bibfield  {author} {\bibinfo {author} {\bibfnamefont {D.}~\bibnamefont
  {Di~Sante}}, \bibinfo {author} {\bibfnamefont {A.}~\bibnamefont {Stroppa}},
  \bibinfo {author} {\bibfnamefont {P.}~\bibnamefont {Barone}}, \bibinfo
  {author} {\bibfnamefont {M.-H.}\ \bibnamefont {Whangbo}},\ and\ \bibinfo
  {author} {\bibfnamefont {S.}~\bibnamefont {Picozzi}},\ }\bibfield  {title}
  {\bibinfo {title} {Emergence of ferroelectricity and spin-valley properties
  in two-dimensional honeycomb binary compounds},\ }\href
  {https://doi.org/10.1103/PhysRevB.91.161401} {\bibfield  {journal} {\bibinfo
  {journal} {Phys. Rev. B}\ }\textbf {\bibinfo {volume} {91}},\ \bibinfo
  {pages} {161401} (\bibinfo {year} {2015})}\BibitemShut {NoStop}%
\bibitem [{\citenamefont {\ifmmode~\mbox{\c{S}}\else \c{S}\fi{}ahin}\ \emph
  {et~al.}(2009)\citenamefont {\ifmmode~\mbox{\c{S}}\else \c{S}\fi{}ahin},
  \citenamefont {Cahangirov}, \citenamefont {Topsakal}, \citenamefont
  {Bekaroglu}, \citenamefont {Akturk}, \citenamefont {Senger},\ and\
  \citenamefont {Ciraci}}]{sahin_monolayer_2009}%
  \BibitemOpen
  \bibfield  {author} {\bibinfo {author} {\bibfnamefont {H.}~\bibnamefont
  {\ifmmode~\mbox{\c{S}}\else \c{S}\fi{}ahin}}, \bibinfo {author}
  {\bibfnamefont {S.}~\bibnamefont {Cahangirov}}, \bibinfo {author}
  {\bibfnamefont {M.}~\bibnamefont {Topsakal}}, \bibinfo {author}
  {\bibfnamefont {E.}~\bibnamefont {Bekaroglu}}, \bibinfo {author}
  {\bibfnamefont {E.}~\bibnamefont {Akturk}}, \bibinfo {author} {\bibfnamefont
  {R.~T.}\ \bibnamefont {Senger}},\ and\ \bibinfo {author} {\bibfnamefont
  {S.}~\bibnamefont {Ciraci}},\ }\bibfield  {title} {\bibinfo {title}
  {Monolayer honeycomb structures of group-iv elements and iii-v binary
  compounds: First-principles calculations},\ }\href
  {https://doi.org/10.1103/PhysRevB.80.155453} {\bibfield  {journal} {\bibinfo
  {journal} {Phys. Rev. B}\ }\textbf {\bibinfo {volume} {80}},\ \bibinfo
  {pages} {155453} (\bibinfo {year} {2009})}\BibitemShut {NoStop}%
\bibitem [{\citenamefont {Chen}\ \emph {et~al.}(2020)\citenamefont {Chen},
  \citenamefont {Liu}, \citenamefont {Zeng}, \citenamefont {Lu}, \citenamefont
  {Lv}, \citenamefont {Chang}, \citenamefont {Lan}, \citenamefont {Wei},
  \citenamefont {Sun}, \citenamefont {Gao}, \citenamefont {Wang},\ and\
  \citenamefont {Fu}}]{Chen2020UniversalGO}%
  \BibitemOpen
  \bibfield  {author} {\bibinfo {author} {\bibfnamefont {Y.}~\bibnamefont
  {Chen}}, \bibinfo {author} {\bibfnamefont {J.}~\bibnamefont {Liu}}, \bibinfo
  {author} {\bibfnamefont {M.}~\bibnamefont {Zeng}}, \bibinfo {author}
  {\bibfnamefont {F.}~\bibnamefont {Lu}}, \bibinfo {author} {\bibfnamefont
  {T.}~\bibnamefont {Lv}}, \bibinfo {author} {\bibfnamefont {Y.}~\bibnamefont
  {Chang}}, \bibinfo {author} {\bibfnamefont {H.}~\bibnamefont {Lan}}, \bibinfo
  {author} {\bibfnamefont {B.}~\bibnamefont {Wei}}, \bibinfo {author}
  {\bibfnamefont {R.}~\bibnamefont {Sun}}, \bibinfo {author} {\bibfnamefont
  {J.}~\bibnamefont {Gao}}, \bibinfo {author} {\bibfnamefont {Z.}~\bibnamefont
  {Wang}},\ and\ \bibinfo {author} {\bibfnamefont {L.}~\bibnamefont {Fu}},\
  }\bibfield  {title} {\bibinfo {title} {Universal growth of ultra-thin iii--v
  semiconductor single crystals},\ }\href
  {https://doi.org/10.1038/s41467-020-17693-5} {\bibfield  {journal} {\bibinfo
  {journal} {Nature Communications}\ }\textbf {\bibinfo {volume} {11}},\
  \bibinfo {pages} {3979} (\bibinfo {year} {2020})}\BibitemShut {NoStop}%
\bibitem [{\citenamefont {Soleimani}\ and\ \citenamefont
  {Pourfath}(2020)}]{Val7}%
  \BibitemOpen
  \bibfield  {author} {\bibinfo {author} {\bibfnamefont {M.}~\bibnamefont
  {Soleimani}}\ and\ \bibinfo {author} {\bibfnamefont {M.}~\bibnamefont
  {Pourfath}},\ }\bibfield  {title} {\bibinfo {title} {Ferroelectricity and
  phase transitions in {In}$_{2}${S}e$_{3}$ van der waals material},\ }\href
  {https://doi.org/10.1039/D0NR04096G} {\bibfield  {journal} {\bibinfo
  {journal} {Nanoscale}\ }\textbf {\bibinfo {volume} {12}},\ \bibinfo {pages}
  {22688} (\bibinfo {year} {2020})}\BibitemShut {NoStop}%
\bibitem [{\citenamefont {Xiao}\ \emph
  {et~al.}(2018{\natexlab{b}})\citenamefont {Xiao}, \citenamefont {Wang},
  \citenamefont {Yang}, \citenamefont {Lu}, \citenamefont {Feng},\ and\
  \citenamefont {Zhang}}]{Curie_2}%
  \BibitemOpen
  \bibfield  {author} {\bibinfo {author} {\bibfnamefont {C.}~\bibnamefont
  {Xiao}}, \bibinfo {author} {\bibfnamefont {F.}~\bibnamefont {Wang}}, \bibinfo
  {author} {\bibfnamefont {S.~A.}\ \bibnamefont {Yang}}, \bibinfo {author}
  {\bibfnamefont {Y.}~\bibnamefont {Lu}}, \bibinfo {author} {\bibfnamefont
  {Y.}~\bibnamefont {Feng}},\ and\ \bibinfo {author} {\bibfnamefont
  {S.}~\bibnamefont {Zhang}},\ }\bibfield  {title} {\bibinfo {title} {Elemental
  ferroelectricity and antiferroelectricity in group-{V} monolayer},\ }\href
  {https://doi.org/https://doi.org/10.1002/adfm.201707383} {\bibfield
  {journal} {\bibinfo  {journal} {Advanced Functional Materials}\ }\textbf
  {\bibinfo {volume} {28}},\ \bibinfo {pages} {1707383} (\bibinfo {year}
  {2018}{\natexlab{b}})}\BibitemShut {NoStop}%
\bibitem [{\citenamefont {Cowley}(1980)}]{LG_1}%
  \BibitemOpen
  \bibfield  {author} {\bibinfo {author} {\bibfnamefont {R.}~\bibnamefont
  {Cowley}},\ }\bibfield  {title} {\bibinfo {title} {Structural phase
  transitions i. landau theory},\ }\href
  {https://doi.org/10.1080/00018738000101346} {\bibfield  {journal} {\bibinfo
  {journal} {Advances in Physics}\ }\textbf {\bibinfo {volume} {29}},\ \bibinfo
  {pages} {1} (\bibinfo {year} {1980})}\BibitemShut {NoStop}%
\bibitem [{\citenamefont {Fei}\ \emph {et~al.}(2016)\citenamefont {Fei},
  \citenamefont {Kang},\ and\ \citenamefont {Yang}}]{Curie_1}%
  \BibitemOpen
  \bibfield  {author} {\bibinfo {author} {\bibfnamefont {R.}~\bibnamefont
  {Fei}}, \bibinfo {author} {\bibfnamefont {W.}~\bibnamefont {Kang}},\ and\
  \bibinfo {author} {\bibfnamefont {L.}~\bibnamefont {Yang}},\ }\bibfield
  {title} {\bibinfo {title} {Ferroelectricity and phase transitions in
  monolayer group-{IV} monochalcogenides},\ }\href
  {https://doi.org/10.1103/PhysRevLett.117.097601} {\bibfield  {journal}
  {\bibinfo  {journal} {Phys. Rev. Lett.}\ }\textbf {\bibinfo {volume} {117}},\
  \bibinfo {pages} {097601} (\bibinfo {year} {2016})}\BibitemShut {NoStop}%
\bibitem [{\citenamefont {Barraza-Lopez}\ \emph {et~al.}(2018)\citenamefont
  {Barraza-Lopez}, \citenamefont {Kaloni}, \citenamefont {Poudel},\ and\
  \citenamefont {Kumar}}]{PhysRevB.97.024110}%
  \BibitemOpen
  \bibfield  {author} {\bibinfo {author} {\bibfnamefont {S.}~\bibnamefont
  {Barraza-Lopez}}, \bibinfo {author} {\bibfnamefont {T.~P.}\ \bibnamefont
  {Kaloni}}, \bibinfo {author} {\bibfnamefont {S.~P.}\ \bibnamefont {Poudel}},\
  and\ \bibinfo {author} {\bibfnamefont {P.}~\bibnamefont {Kumar}},\ }\bibfield
   {title} {\bibinfo {title} {Tuning the ferroelectric-to-paraelectric
  transition temperature and dipole orientation of group-iv monochalcogenide
  monolayers},\ }\href {https://doi.org/10.1103/PhysRevB.97.024110} {\bibfield
  {journal} {\bibinfo  {journal} {Phys. Rev. B}\ }\textbf {\bibinfo {volume}
  {97}},\ \bibinfo {pages} {024110} (\bibinfo {year} {2018})}\BibitemShut
  {NoStop}%
\bibitem [{\citenamefont {Liu}\ \emph {et~al.}(2018)\citenamefont {Liu},
  \citenamefont {Wan}, \citenamefont {Ma}, \citenamefont {Guo},\ and\
  \citenamefont {Yao}}]{C7NR09006D}%
  \BibitemOpen
  \bibfield  {author} {\bibinfo {author} {\bibfnamefont {C.}~\bibnamefont
  {Liu}}, \bibinfo {author} {\bibfnamefont {W.}~\bibnamefont {Wan}}, \bibinfo
  {author} {\bibfnamefont {J.}~\bibnamefont {Ma}}, \bibinfo {author}
  {\bibfnamefont {W.}~\bibnamefont {Guo}},\ and\ \bibinfo {author}
  {\bibfnamefont {Y.}~\bibnamefont {Yao}},\ }\bibfield  {title} {\bibinfo
  {title} {Robust ferroelectricity in two-dimensional sbn and bip},\ }\href
  {https://doi.org/10.1039/C7NR09006D} {\bibfield  {journal} {\bibinfo
  {journal} {Nanoscale}\ }\textbf {\bibinfo {volume} {10}},\ \bibinfo {pages}
  {7984} (\bibinfo {year} {2018})}\BibitemShut {NoStop}%
\bibitem [{\citenamefont {Wang}\ \emph {et~al.}(2021)\citenamefont {Wang},
  \citenamefont {Qiao},\ and\ \citenamefont {Kim}}]{Wang_HfO2}%
  \BibitemOpen
  \bibfield  {author} {\bibinfo {author} {\bibfnamefont {C.}~\bibnamefont
  {Wang}}, \bibinfo {author} {\bibfnamefont {H.}~\bibnamefont {Qiao}},\ and\
  \bibinfo {author} {\bibfnamefont {Y.}~\bibnamefont {Kim}},\ }\bibfield
  {title} {\bibinfo {title} {{Perspective on the switching behavior of
  HfO2-based ferroelectrics}},\ }\href {https://doi.org/10.1063/5.0035652}
  {\bibfield  {journal} {\bibinfo  {journal} {Journal of Applied Physics}\
  }\textbf {\bibinfo {volume} {129}},\ \bibinfo {pages} {010902} (\bibinfo
  {year} {2021})}\BibitemShut {NoStop}%
\bibitem [{\citenamefont {Sokolov}\ \emph {et~al.}(2015)\citenamefont
  {Sokolov}, \citenamefont {Bak}, \citenamefont {Lu}, \citenamefont {Li},
  \citenamefont {Tsymbal},\ and\ \citenamefont {Gruverman}}]{TER_form_1}%
  \BibitemOpen
  \bibfield  {author} {\bibinfo {author} {\bibfnamefont {A.}~\bibnamefont
  {Sokolov}}, \bibinfo {author} {\bibfnamefont {O.}~\bibnamefont {Bak}},
  \bibinfo {author} {\bibfnamefont {H.}~\bibnamefont {Lu}}, \bibinfo {author}
  {\bibfnamefont {S.}~\bibnamefont {Li}}, \bibinfo {author} {\bibfnamefont
  {E.~Y.}\ \bibnamefont {Tsymbal}},\ and\ \bibinfo {author} {\bibfnamefont
  {A.}~\bibnamefont {Gruverman}},\ }\bibfield  {title} {\bibinfo {title}
  {Effect of epitaxial strain on tunneling electroresistance in ferroelectric
  tunnel junctions},\ }\href {https://doi.org/10.1088/0957-4484/26/30/305202}
  {\bibfield  {journal} {\bibinfo  {journal} {Nanotechnology}\ }\textbf
  {\bibinfo {volume} {26}},\ \bibinfo {pages} {305202} (\bibinfo {year}
  {2015})}\BibitemShut {NoStop}%
\bibitem [{\citenamefont {Gruverman}\ \emph {et~al.}(2009)\citenamefont
  {Gruverman}, \citenamefont {Wu}, \citenamefont {Lu}, \citenamefont {Wang},
  \citenamefont {Jang}, \citenamefont {Folkman}, \citenamefont {Zhuravlev},
  \citenamefont {Felker}, \citenamefont {Rzchowski}, \citenamefont {Eom},\ and\
  \citenamefont {Tsymbal}}]{TER_form_2}%
  \BibitemOpen
  \bibfield  {author} {\bibinfo {author} {\bibfnamefont {A.}~\bibnamefont
  {Gruverman}}, \bibinfo {author} {\bibfnamefont {D.}~\bibnamefont {Wu}},
  \bibinfo {author} {\bibfnamefont {H.}~\bibnamefont {Lu}}, \bibinfo {author}
  {\bibfnamefont {Y.}~\bibnamefont {Wang}}, \bibinfo {author} {\bibfnamefont
  {H.~W.}\ \bibnamefont {Jang}}, \bibinfo {author} {\bibfnamefont {C.~M.}\
  \bibnamefont {Folkman}}, \bibinfo {author} {\bibfnamefont {M.~Y.}\
  \bibnamefont {Zhuravlev}}, \bibinfo {author} {\bibfnamefont {D.}~\bibnamefont
  {Felker}}, \bibinfo {author} {\bibfnamefont {M.}~\bibnamefont {Rzchowski}},
  \bibinfo {author} {\bibfnamefont {C.-B.}\ \bibnamefont {Eom}},\ and\ \bibinfo
  {author} {\bibfnamefont {E.~Y.}\ \bibnamefont {Tsymbal}},\ }\bibfield
  {title} {\bibinfo {title} {Tunneling electroresistance effect in
  ferroelectric tunnel junctions at the nanoscale},\ }\href
  {https://doi.org/10.1021/nl901754t} {\bibfield  {journal} {\bibinfo
  {journal} {Nano Letters}\ }\textbf {\bibinfo {volume} {9}},\ \bibinfo {pages}
  {3539} (\bibinfo {year} {2009})}\BibitemShut {NoStop}%
\bibitem [{\citenamefont {Velev}\ \emph {et~al.}(2016)\citenamefont {Velev},
  \citenamefont {Burton}, \citenamefont {Zhuravlev},\ and\ \citenamefont
  {Tsymbal}}]{TER_form_3}%
  \BibitemOpen
  \bibfield  {author} {\bibinfo {author} {\bibfnamefont {J.~P.}\ \bibnamefont
  {Velev}}, \bibinfo {author} {\bibfnamefont {J.~D.}\ \bibnamefont {Burton}},
  \bibinfo {author} {\bibfnamefont {M.~Y.}\ \bibnamefont {Zhuravlev}},\ and\
  \bibinfo {author} {\bibfnamefont {E.~Y.}\ \bibnamefont {Tsymbal}},\
  }\bibfield  {title} {\bibinfo {title} {Predictive modelling of ferroelectric
  tunnel junctions},\ }\href {https://doi.org/10.1038/npjcompumats.2016.9}
  {\bibfield  {journal} {\bibinfo  {journal} {npj Computational Materials}\
  }\textbf {\bibinfo {volume} {2}},\ \bibinfo {pages} {16009} (\bibinfo {year}
  {2016})}\BibitemShut {NoStop}%
\bibitem [{\citenamefont {Lahiri}\ \emph {et~al.}(2022)\citenamefont {Lahiri},
  \citenamefont {Bhore}, \citenamefont {Das},\ and\ \citenamefont
  {Agarwal}}]{Lahiri_22}%
  \BibitemOpen
  \bibfield  {author} {\bibinfo {author} {\bibfnamefont {S.}~\bibnamefont
  {Lahiri}}, \bibinfo {author} {\bibfnamefont {T.}~\bibnamefont {Bhore}},
  \bibinfo {author} {\bibfnamefont {K.}~\bibnamefont {Das}},\ and\ \bibinfo
  {author} {\bibfnamefont {A.}~\bibnamefont {Agarwal}},\ }\bibfield  {title}
  {\bibinfo {title} {Nonlinear magnetoresistivity in two-dimensional systems
  induced by berry curvature},\ }\href
  {https://doi.org/10.1103/PhysRevB.105.045421} {\bibfield  {journal} {\bibinfo
   {journal} {Phys. Rev. B}\ }\textbf {\bibinfo {volume} {105}},\ \bibinfo
  {pages} {045421} (\bibinfo {year} {2022})}\BibitemShut {NoStop}%
\bibitem [{\citenamefont {Bhalla}\ \emph {et~al.}(2022)\citenamefont {Bhalla},
  \citenamefont {Das}, \citenamefont {Culcer},\ and\ \citenamefont
  {Agarwal}}]{PhysRevLett.129.227401}%
  \BibitemOpen
  \bibfield  {author} {\bibinfo {author} {\bibfnamefont {P.}~\bibnamefont
  {Bhalla}}, \bibinfo {author} {\bibfnamefont {K.}~\bibnamefont {Das}},
  \bibinfo {author} {\bibfnamefont {D.}~\bibnamefont {Culcer}},\ and\ \bibinfo
  {author} {\bibfnamefont {A.}~\bibnamefont {Agarwal}},\ }\bibfield  {title}
  {\bibinfo {title} {Resonant second-harmonic generation as a probe of quantum
  geometry},\ }\href {https://doi.org/10.1103/PhysRevLett.129.227401}
  {\bibfield  {journal} {\bibinfo  {journal} {Phys. Rev. Lett.}\ }\textbf
  {\bibinfo {volume} {129}},\ \bibinfo {pages} {227401} (\bibinfo {year}
  {2022})}\BibitemShut {NoStop}%
\bibitem [{\citenamefont {Bhalla}\ \emph {et~al.}(2023)\citenamefont {Bhalla},
  \citenamefont {Das}, \citenamefont {Agarwal},\ and\ \citenamefont
  {Culcer}}]{PhysRevB.107.165131}%
  \BibitemOpen
  \bibfield  {author} {\bibinfo {author} {\bibfnamefont {P.}~\bibnamefont
  {Bhalla}}, \bibinfo {author} {\bibfnamefont {K.}~\bibnamefont {Das}},
  \bibinfo {author} {\bibfnamefont {A.}~\bibnamefont {Agarwal}},\ and\ \bibinfo
  {author} {\bibfnamefont {D.}~\bibnamefont {Culcer}},\ }\bibfield  {title}
  {\bibinfo {title} {Quantum kinetic theory of nonlinear optical currents:
  Finite fermi surface and fermi sea contributions},\ }\href
  {https://doi.org/10.1103/PhysRevB.107.165131} {\bibfield  {journal} {\bibinfo
   {journal} {Phys. Rev. B}\ }\textbf {\bibinfo {volume} {107}},\ \bibinfo
  {pages} {165131} (\bibinfo {year} {2023})}\BibitemShut {NoStop}%
\bibitem [{\citenamefont {Das}\ \emph {et~al.}(2023)\citenamefont {Das},
  \citenamefont {Lahiri}, \citenamefont {Atencia}, \citenamefont {Culcer},\
  and\ \citenamefont {Agarwal}}]{PhysRevB.108.L201405}%
  \BibitemOpen
  \bibfield  {author} {\bibinfo {author} {\bibfnamefont {K.}~\bibnamefont
  {Das}}, \bibinfo {author} {\bibfnamefont {S.}~\bibnamefont {Lahiri}},
  \bibinfo {author} {\bibfnamefont {R.~B.}\ \bibnamefont {Atencia}}, \bibinfo
  {author} {\bibfnamefont {D.}~\bibnamefont {Culcer}},\ and\ \bibinfo {author}
  {\bibfnamefont {A.}~\bibnamefont {Agarwal}},\ }\bibfield  {title} {\bibinfo
  {title} {Intrinsic nonlinear conductivities induced by the quantum metric},\
  }\href {https://doi.org/10.1103/PhysRevB.108.L201405} {\bibfield  {journal}
  {\bibinfo  {journal} {Phys. Rev. B}\ }\textbf {\bibinfo {volume} {108}},\
  \bibinfo {pages} {L201405} (\bibinfo {year} {2023})}\BibitemShut {NoStop}%
\bibitem [{\citenamefont {Zhang}\ and\ \citenamefont
  {Fu}(2021)}]{zhang2021terahertz}%
  \BibitemOpen
  \bibfield  {author} {\bibinfo {author} {\bibfnamefont {Y.}~\bibnamefont
  {Zhang}}\ and\ \bibinfo {author} {\bibfnamefont {L.}~\bibnamefont {Fu}},\
  }\bibfield  {title} {\bibinfo {title} {Terahertz detection based on nonlinear
  hall effect without magnetic field},\ }\href
  {https://doi.org/10.1073/pnas.2100736118} {\bibfield  {journal} {\bibinfo
  {journal} {Proceedings of the National Academy of Sciences}\ }\textbf
  {\bibinfo {volume} {118}},\ \bibinfo {pages} {e2100736118} (\bibinfo {year}
  {2021})}\BibitemShut {NoStop}%
\bibitem [{\citenamefont {Perdew}\ \emph {et~al.}(1996)\citenamefont {Perdew},
  \citenamefont {Burke},\ and\ \citenamefont
  {Ernzerhof}}]{PhysRevLett.77.3865}%
  \BibitemOpen
  \bibfield  {author} {\bibinfo {author} {\bibfnamefont {J.~P.}\ \bibnamefont
  {Perdew}}, \bibinfo {author} {\bibfnamefont {K.}~\bibnamefont {Burke}},\ and\
  \bibinfo {author} {\bibfnamefont {M.}~\bibnamefont {Ernzerhof}},\ }\bibfield
  {title} {\bibinfo {title} {Generalized gradient approximation made simple},\
  }\href {https://doi.org/10.1103/PhysRevLett.77.3865} {\bibfield  {journal}
  {\bibinfo  {journal} {Phys. Rev. Lett.}\ }\textbf {\bibinfo {volume} {77}},\
  \bibinfo {pages} {3865} (\bibinfo {year} {1996})}\BibitemShut {NoStop}%
\bibitem [{\citenamefont {Kresse}\ and\ \citenamefont
  {Furthm\"uller}(1996)}]{PhysRevB.54.11169}%
  \BibitemOpen
  \bibfield  {author} {\bibinfo {author} {\bibfnamefont {G.}~\bibnamefont
  {Kresse}}\ and\ \bibinfo {author} {\bibfnamefont {J.}~\bibnamefont
  {Furthm\"uller}},\ }\bibfield  {title} {\bibinfo {title} {Efficient iterative
  schemes for ab initio total-energy calculations using a plane-wave basis
  set},\ }\href {https://doi.org/10.1103/PhysRevB.54.11169} {\bibfield
  {journal} {\bibinfo  {journal} {Phys. Rev. B}\ }\textbf {\bibinfo {volume}
  {54}},\ \bibinfo {pages} {11169} (\bibinfo {year} {1996})}\BibitemShut
  {NoStop}%
\bibitem [{\citenamefont {Pizzi}\ \emph {et~al.}(2020)\citenamefont {Pizzi},
  \citenamefont {Vitale}, \citenamefont {Arita}, \citenamefont {Blügel},
  \citenamefont {Freimuth}, \citenamefont {Géranton}, \citenamefont
  {Gibertini}, \citenamefont {Gresch}, \citenamefont {Johnson}, \citenamefont
  {Koretsune}, \citenamefont {Ibañez-Azpiroz}, \citenamefont {Lee},
  \citenamefont {Lihm}, \citenamefont {Marchand}, \citenamefont {Marrazzo},
  \citenamefont {Mokrousov}, \citenamefont {Mustafa}, \citenamefont {Nohara},
  \citenamefont {Nomura}, \citenamefont {Paulatto}, \citenamefont {Poncé},
  \citenamefont {Ponweiser}, \citenamefont {Qiao}, \citenamefont {Thöle},
  \citenamefont {Tsirkin}, \citenamefont {Wierzbowska}, \citenamefont
  {Marzari}, \citenamefont {Vanderbilt}, \citenamefont {Souza}, \citenamefont
  {Mostofi},\ and\ \citenamefont {Yates}}]{pizzi2020wannier90}%
  \BibitemOpen
  \bibfield  {author} {\bibinfo {author} {\bibfnamefont {G.}~\bibnamefont
  {Pizzi}}, \bibinfo {author} {\bibfnamefont {V.}~\bibnamefont {Vitale}},
  \bibinfo {author} {\bibfnamefont {R.}~\bibnamefont {Arita}}, \bibinfo
  {author} {\bibfnamefont {S.}~\bibnamefont {Blügel}}, \bibinfo {author}
  {\bibfnamefont {F.}~\bibnamefont {Freimuth}}, \bibinfo {author}
  {\bibfnamefont {G.}~\bibnamefont {Géranton}}, \bibinfo {author}
  {\bibfnamefont {M.}~\bibnamefont {Gibertini}}, \bibinfo {author}
  {\bibfnamefont {D.}~\bibnamefont {Gresch}}, \bibinfo {author} {\bibfnamefont
  {C.}~\bibnamefont {Johnson}}, \bibinfo {author} {\bibfnamefont
  {T.}~\bibnamefont {Koretsune}}, \bibinfo {author} {\bibfnamefont
  {J.}~\bibnamefont {Ibañez-Azpiroz}}, \bibinfo {author} {\bibfnamefont
  {H.}~\bibnamefont {Lee}}, \bibinfo {author} {\bibfnamefont {J.-M.}\
  \bibnamefont {Lihm}}, \bibinfo {author} {\bibfnamefont {D.}~\bibnamefont
  {Marchand}}, \bibinfo {author} {\bibfnamefont {A.}~\bibnamefont {Marrazzo}},
  \bibinfo {author} {\bibfnamefont {Y.}~\bibnamefont {Mokrousov}}, \bibinfo
  {author} {\bibfnamefont {J.~I.}\ \bibnamefont {Mustafa}}, \bibinfo {author}
  {\bibfnamefont {Y.}~\bibnamefont {Nohara}}, \bibinfo {author} {\bibfnamefont
  {Y.}~\bibnamefont {Nomura}}, \bibinfo {author} {\bibfnamefont
  {L.}~\bibnamefont {Paulatto}}, \bibinfo {author} {\bibfnamefont
  {S.}~\bibnamefont {Poncé}}, \bibinfo {author} {\bibfnamefont
  {T.}~\bibnamefont {Ponweiser}}, \bibinfo {author} {\bibfnamefont
  {J.}~\bibnamefont {Qiao}}, \bibinfo {author} {\bibfnamefont {F.}~\bibnamefont
  {Thöle}}, \bibinfo {author} {\bibfnamefont {S.~S.}\ \bibnamefont {Tsirkin}},
  \bibinfo {author} {\bibfnamefont {M.}~\bibnamefont {Wierzbowska}}, \bibinfo
  {author} {\bibfnamefont {N.}~\bibnamefont {Marzari}}, \bibinfo {author}
  {\bibfnamefont {D.}~\bibnamefont {Vanderbilt}}, \bibinfo {author}
  {\bibfnamefont {I.}~\bibnamefont {Souza}}, \bibinfo {author} {\bibfnamefont
  {A.~A.}\ \bibnamefont {Mostofi}},\ and\ \bibinfo {author} {\bibfnamefont
  {J.~R.}\ \bibnamefont {Yates}},\ }\bibfield  {title} {\bibinfo {title}
  {Wannier90 as a community code: new features and applications},\ }\href
  {https://doi.org/10.1088/1361-648X/ab51ff} {\bibfield  {journal} {\bibinfo
  {journal} {Journal of Physics: Condensed Matter}\ }\textbf {\bibinfo {volume}
  {32}},\ \bibinfo {pages} {165902} (\bibinfo {year} {2020})}\BibitemShut
  {NoStop}%
\bibitem [{\citenamefont {You}\ \emph {et~al.}(2018)\citenamefont {You},
  \citenamefont {Fang}, \citenamefont {Xu}, \citenamefont {Kaxiras},\ and\
  \citenamefont {Low}}]{PhysRevB.98.121109}%
  \BibitemOpen
  \bibfield  {author} {\bibinfo {author} {\bibfnamefont {J.-S.}\ \bibnamefont
  {You}}, \bibinfo {author} {\bibfnamefont {S.}~\bibnamefont {Fang}}, \bibinfo
  {author} {\bibfnamefont {S.-Y.}\ \bibnamefont {Xu}}, \bibinfo {author}
  {\bibfnamefont {E.}~\bibnamefont {Kaxiras}},\ and\ \bibinfo {author}
  {\bibfnamefont {T.}~\bibnamefont {Low}},\ }\bibfield  {title} {\bibinfo
  {title} {Berry curvature dipole current in the transition metal
  dichalcogenides family},\ }\href {https://doi.org/10.1103/PhysRevB.98.121109}
  {\bibfield  {journal} {\bibinfo  {journal} {Phys. Rev. B}\ }\textbf {\bibinfo
  {volume} {98}},\ \bibinfo {pages} {121109} (\bibinfo {year}
  {2018})}\BibitemShut {NoStop}%
\bibitem [{\citenamefont {Zhang}\ \emph {et~al.}(2018)\citenamefont {Zhang},
  \citenamefont {van~den Brink}, \citenamefont {Felser},\ and\ \citenamefont
  {Yan}}]{zhang2018electrically}%
  \BibitemOpen
  \bibfield  {author} {\bibinfo {author} {\bibfnamefont {Y.}~\bibnamefont
  {Zhang}}, \bibinfo {author} {\bibfnamefont {J.}~\bibnamefont {van~den
  Brink}}, \bibinfo {author} {\bibfnamefont {C.}~\bibnamefont {Felser}},\ and\
  \bibinfo {author} {\bibfnamefont {B.}~\bibnamefont {Yan}},\ }\bibfield
  {title} {\bibinfo {title} {Electrically tuneable nonlinear anomalous hall
  effect in two-dimensional transition-metal dichalcogenides {WTe}$_2$ and
  {MoTe}$_2$},\ }\href {https://doi.org/10.1088/2053-1583/aad1ae} {\bibfield
  {journal} {\bibinfo  {journal} {2D Materials}\ }\textbf {\bibinfo {volume}
  {5}},\ \bibinfo {pages} {044001} (\bibinfo {year} {2018})}\BibitemShut
  {NoStop}%
\end{thebibliography}%
\end{document}